\documentclass[11pt]{article}

\usepackage[final]{acl}

\usepackage{times}
\usepackage{latexsym}

\usepackage[T1]{fontenc}

\usepackage[utf8]{inputenc}

\usepackage{microtype}

\usepackage{inconsolata}

\usepackage{graphicx}

\usepackage{algorithm}
\usepackage{algorithmicx}
\usepackage{multirow}
\usepackage[noend]{algpseudocode}
\usepackage{enumitem}
\usepackage{threeparttable}
\usepackage{booktabs}
\usepackage{amsmath}
\usepackage{float}

\title{MERIT: Mitigating Exposure Bias in Generative XMC for User-Interest Propensity Modeling}

\author{Abhinav Mahajan \\
  Carnegie Mellon University \\
  Pittsburgh, PA, USA \\
  \texttt{abhinavm@andrew.cmu.edu} \\\And
  Arindam Sarkar \\
  Amazon \\
  Bangalore, India \\
  \texttt{arindsar@amazon.com} \\\And
  Prakash Mandayam Comar \\
  Amazon \\
  Bangalore, India \\
  \texttt{prakasc@amazon.com} \\ }

\begin{document}
\maketitle
\begin{abstract}

Matching users to interest categories at scale is central to personalized shopping, but the task is challenging in large e-commerce platforms, where label spaces continually evolve and user-interest signals are sparse and long-tailed.
Autoregressive language models are appealing because their world 
knowledge and semantic priors over descriptors generalize across 
extreme label spaces and accommodate multiple valid label assignments.
Yet under teacher-forced fine-tuning, inference-time predictions become part of the conditioning context: early errors steer later outputs toward co-occurring labels, over-generating near-correlates and missing unrelated true interests. We present \textbf{MERIT}, a framework for user-interest propensity modeling that mitigates this \textit{exposure bias} through a self-correction objective. A permutation-invariant multi-target loss over 
shuffled mixtures of gold and mined hard-negative labels
exposes the generator to erroneous prefixes while preserving the efficiency of 
teacher-forced training. 
This training objective concentrates supervision at classification positions, yielding propensity-aligned hidden states
powering a lightweight scorer for bidirectional retrieval (interests for users and users for interests). 
On proprietary e-commerce dataset with 250k+ interest categories, 
MERIT improves global recall by at least \textbf{11.9\%} and average Hit@k by  \textbf{6.1\%}.
In production A/B tests, it achieves \textbf{+0.26\% } 
gain in user conversion.

\end{abstract}

\section{Introduction}
\label{sec:intro}
A key problem in personalization is predicting a
user's propensity to engage with a candidate entity such as a product,
collection, or campaign. Such candidates are highly
dynamic~\cite{10.1145/3705328.3748112} (continually launched and
retired) creating cold-start regimes where candidate-level learning is
brittle due to lack of historical signal~\cite{LIKA20142065,DBLP:journals/corr/abs-2501-01945}. 
A robust alternative 
is to employ \emph{semantic interest
descriptors} that stay stable across content churn, which recent
generative tagging~\cite{10.1145/3711896.3737242,Li2023TagGPTLL,10.1007/978-3-031-78090-5_4}
makes feasible at scale: a model assigns interpretable descriptors
(e.g., ``running'', ``home organization'') to entities from their text
and metadata, shifting the learning target from each ephemeral candidate
to a stable space that generalizes to new candidates via their tags.
This is attractive at deployment, since the binding constraint on the targeting layer is training cost. 
Typical discriminative one-versus-all propensity models over the descriptor vocabulary scale linearly in the number of descriptors, restricting deployment to high-traffic categories and precluding vocabulary evolution without manual retraining.

This motivates \emph{candidate-free} interest propensity modeling at
extreme label scale, based on a user's interaction history: predict a
sparse set of descriptors from a large vocabulary and support
bidirectional retrieval, ranking interests for a user
(personalization) and users for an interest (segmentation). The setting
has the hallmarks of extreme multi-label classification (XMC), i.e., a
long-tailed label space, sparse positives, and ambiguous
non-interactions (unobserved interests are not true negatives) with
two consequences. Typical e-commerce users hold multiple, often unrelated interests,
so the goal is to recover a \emph{diverse} interest set rather than
variants of one. We treat this as a retrieval problem since the predicted interests are not a final user-facing decision but an upstream retrieval interface, refined downstream by business rules and eligibility constraints. As in any retrieval setting,
downstream filtering is cheap and missed candidates are irrecoverable, recall is therefore the operative metric here.

Generative language models are a natural fit here, as they operate directly over textual descriptors, generalize across long-tailed spaces, and produce structured multi-label outputs that are auditable. LM finetuning further transfers semantic priors over descriptors and entities to 
user-interest prediction. 
This is useful because descriptor co-occurrence is a tractable signal in semantic space: 
rather than memorizing 250k+ categories independently, the model can exploit structure among them. But this reliance on correlation is double-edged. With sparse per-descriptor supervision, the model must use co-occurrence to make predictions, but must also override it when the user history supports unrelated interests.
Exposure bias sharpens this tension. Teacher forcing conditions on gold prefixes, while inference conditions on the model’s own outputs~\cite{ranzato2015sequence,zhang-etal-2019-bridging}. In fluent-text generation, this mismatch often appears as degraded sequence quality; in multi-label generation, it becomes a recall failure.
Once decoding commits to a semantic region (e.g., “sports”), the model can over-amplify co-occurrence and emit near-correlates (“athletic gear”, “fitness equipment”) in a self-reinforcing cascade, suppressing unrelated true interests.
These cascades are not incidental: training data over-represents certain descriptor co-occurrences, and pretrained models bring their own priors. They also collapse representations toward the dominant region, mirroring the single-vector bottleneck of dual encoders and obstructing calibrated cross-user propensity scoring.

We introduce \textbf{MERIT}, a framework for user-interest propensity modeling at extreme label scale (250k+ labels). Our key contributions are: (i) We identify exposure bias in generative XMC as a recall failure from co-occurrence cascades, and address it with a permutation-invariant multi-target loss over mixed gold/mined prefixes that trains recovery within a standard teacher forcing pass without per-step generation, improving recall by at least \textbf{11.9\%}, 
(ii) show that training on this objective yields, by design, propensity-aligned hidden states and informative hard negatives that together enable the training of a lightweight scorer, improving average Hit@k by atleast \textbf{6.1\%}, (iii) validate MERIT at production scale via A/B tests, showing positive posterior mean conversion lift averaging +0.26\%.

\section{Background \& Related Work}
\textbf{User interest modeling and XMC retrieval:}
CTR-style user-interest models~\cite{10.1145/3219819.3219823,10.1609/aaai.v33i01.33015941,10.1145/3357384.3357814,qi-etal-2021-hierec} score a \emph{given} candidate against history; we instead \emph{predict} a descriptor set from an extreme vocabulary as a retrieval interface. XMC targets such label spaces via tree-based~\cite{prabhu2018parabel,jain2016xmlcnn}, embedding-based~\cite{dahiya2021deepxml,mittal2021decaf,dahiya2023ngame}, and retrieval-style dual-encoder methods~\cite{10.5555/3454287.3454810,Chang2019TamingPT,you2019attentionxml,jiang2021lightxml,kharbanda2022renee,dhillon2024dual}. Dual encoders are parameter-efficient and generalize to new labels, but hinge on informative hard negatives that are costly to mine at extreme scale. In MERIT, the generator's own confident errors furnish such negatives for free.

\textbf{Generative and code-based XMC:}
Generative XMC predicts labels autoregressively~\cite{sun2022clusterguided,simig2022open}; related generative-retrieval work replaces text labels with discrete Semantic IDs~\cite{10.5555/3666122.3666574,10.5555/3600270.3601857}, which aid latency but sacrifice interpretability, so we retain text-level descriptors. LLM-based recommendation~\cite{10.1145/3604915.3608857,10.1145/3523227.3546767,NEURIPS2023_91228b94,geng2022recommendation,hou2023large,lin2023rella,cui2022m6rec} via prompting lacks calibrated propensities at extreme scale~\cite{pmlr-v139-zhao21c}; we therefore use the LLM to supply training signal, delegating cross-user scoring to a distilled model.

\begin{figure*}[!h]
    \centering
    \includegraphics[width=1.85\columnwidth]{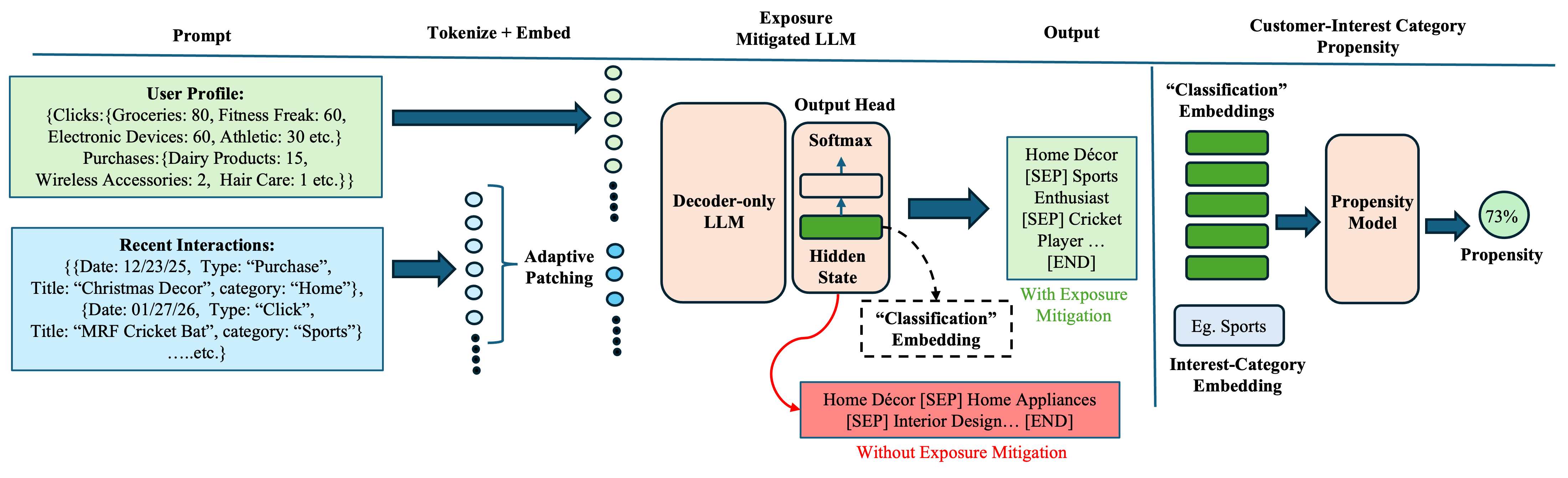}
    \caption{\textbf{MERIT inference pipeline}. User profile (aggregated interest-categories) and recent interactions (raw product data) are tokenized, embedded, and compressed via adaptive patching. The exposure-mitigated LLM autoregressively generates interest-categories. \emph{With mitigation} (green), the model produces diverse predictions (e.g., Sports Enthusiast); \emph{without mitigation} (red), it enters correlation loops (Home Décor → Home Appliances → Interior Design). Hidden states from the first token of the first few
    generated categories are extracted as enriched classification embeddings, which serve as input to the propensity model for scoring any interest category.}
    \label{fig:pipeline}
\end{figure*}

\textbf{Exposure bias in autoregressive generation:}
The gap between teacher-forced training and free-running inference~\cite{ranzato2015sequence,zhang-etal-2019-bridging} has been addressed via scheduled sampling~\cite{bengio2015scheduled}, reinforcement learning~\cite{ranzato2015sequence}, and imitation learning~\cite{ross2011reduction}, largely in fluent generation, where the failure mode is degraded sequence quality. We study the same mismatch in \emph{multi-label} generation, where early label errors can amplify co-occurrence priors into correlation cascades reducing coverage of true label set. MERIT addresses this with an efficient self-correction objective: mixed gold/error prefixes expose the model to its own mistakes and a multi-target teacher-forced loss learns recovery without per-step generation.

\section{Methodology}
\label{sec:method}

We predict user propensities over an 
extremely large interest-category space $\mathcal{L} = \{l_1, \dots, l_K\}$
that acts as a retrieval interface to dynamic entities such as products and campaigns. Each product carries several multi-token category tags (e.g., a cricket bat: "sports enthusiast", "cricket player", "fitness equipment").
For a user $c$ purchasing at time $t_{\text{purchase}}$, we form one example: the \textbf{input} is the interaction history $H_c = \{(p, t)\}$ of products clicked or purchased in the preceding year, excluding the final 7 days ($t < t_{\text{purchase}} - 7\text{d}$); the \textbf{output} is the tag set $Y_c \subset \mathcal{L}$ of the purchased product. The exclusion window prevents trivial prediction from immediate pre-purchase signals and matches production settings where scores need to be pre-computed.

Finetuned LLMs are well-suited for this task, but pure generation at inference yields no calibrated scores comparable \emph{across} users, blocking the interest$\rightarrow$user retrieval needed for segmentation. We address this by training an exposure-mitigated generator (\S\ref{sec:self-correction}) with adaptive patching for long histories (\S\ref{sec:patching}), then training a lightweight propensity model on the generator's hidden states (\S\ref{sec:dual-encoder}), reusing its own errors as hard negatives.

\subsection{Exposure-Mitigation for Self-Correction}
\label{sec:self-correction}

Standard teacher forcing conditions on gold prefixes during training but on the model's own predictions at inference. This exposure bias makes the model exploit prefix correlations rather than classify independently: once an early token commits to a semantic region (e.g., "sports"), later outputs cluster on correlated variants ("athletic gear", "fitness equipment") instead of diverse, unrelated interests, directly harming recall. We mitigate this with a self-correction objective that teaches the model to recover from its own errors, in two steps.

\textbf{Step 1: Offline hard-negative mining.} We first train an initial generator with standard cross-entropy, then run independent temperature sampling over the training set to mine plausible but incorrect labels $\hat{Y}_{\text{filtered}}$, augmenting the golden tag set into the shuffled sequence, $\mathcal{L}_c = Y_c \cup \hat{Y}_{\text{filtered}}$. Because this inference pass is performed offline, it is far cheaper than online sampling methods that generate at every training step. 
We drop any mined negative whose Cohere similarity to a gold label exceeds $0.6$, intentionally below our $\tau{=}0.7$ soft-match evaluation threshold so that near-synonyms and paraphrases of a user's true interests are not injected into the prefix as negatives (see Limitations for a discussion of residual false-negative risk).

\textbf{Step 2: Exposure-mitigated teacher forcing.} For each user we shuffle $\mathcal{L}_c$ and concatenate its labels with a special \texttt{[SEP]} token into a target sequence. We retain the efficiency of teacher forcing by scoring the full sequence (prompt + shuffled target) in a \emph{single forward pass} with no generation; the only change is to the per-token loss, which we modify to admit \emph{multiple valid targets} while never rewarding a mined negative as a label choice, so the model is exposed to incorrect labels in the prefix without learning to reproduce them. 
At each position $t$, we define the \emph{prefix} as the tokens consumed so far within the current label (i.e., since the last \texttt{[SEP]}); the valid set $\mathcal{V}_t$ depends on this prefix:

\begin{itemize}[leftmargin=*,itemsep=0pt,topsep=0pt]
    \item \emph{Empty prefix} (a \emph{classification} position): $\mathcal{V}_t$ is the first tokens of every not-yet-completed golden label, or \texttt{[END]} if none remain. Pooling over these first tokens makes the loss \textbf{permutation-invariant} so that the model is not penalized for ordering golden labels differently.
    \item \emph{Prefix matches golden label(s)}: $\mathcal{V}_t$ is their next tokens, covering shared prefixes (e.g., "sports enthusiast" / "sports gear").
    \item \emph{Prefix matches only a hard negative} (e.g., "kitchen" from "kitchen appliances"): $\mathcal{V}_t$ completes that negative, enforcing \textbf{label coherence} so even incorrect labels stay well-formed rather than switching mid-phrase.
\end{itemize}
Whenever the prefix completes a label, \texttt{[SEP]} joins $\mathcal{V}_t$. The loss pools probability mass over $\mathcal{V}_t$:
\begin{equation}
\ell_t = -\log \left( \frac{\sum_{v \in \mathcal{V}_t} \exp(z_t^{(v)})}{\sum_{v'} \exp(z_t^{(v')})} \right).
\label{eq:selective_loss}
\end{equation}
Algo.~\ref{alg:self-correct} gives the full procedure. We \textbf{strongly encourage} the reader to go through the intricate example dry-run in \textbf{Appendix~\ref{app:dry-run}}.

\begin{algorithm}[h]
\caption{Self-Correction Loss Computation}
\label{alg:self-correct}
\small
\begin{algorithmic}[1]
\Require Golden labels $Y_c$, mined hard negatives $\hat{Y}_{\text{filtered}}$
\State $\mathcal{L}_c \gets Y_c \cup \hat{Y}_{\text{filtered}}$ \Comment{Augmented label set}
\State $L \gets \text{Shuffle}(\mathcal{L}_c)$
\State $S \gets \text{Tokenize}(L \oplus \texttt{[SEP]})$ \Comment{Target sequence}
\State $X \gets \text{prompt} \oplus S$
\State $\{z_t\}_{t=1}^T \gets \text{Model}(X)$ \Comment{Target Logits, single pass}
\State $\text{prefix} \gets []$, $\text{completed} \gets \{\}$
\For{$t = 1$ to $T$}
    \If{$\text{prefix} = []$} \Comment{Starting a new label}
        \State $\mathcal{V}_t \gets$ \{first token of $\ell \mid \ell \in Y_c \setminus \text{completed}\}$
        \If{$Y_c \setminus \text{completed} = \emptyset$}
            \State $\mathcal{V}_t \gets \{\texttt{[END]}\}$
        \EndIf
    \Else \Comment{Continuing a label}
        \State $G \gets \{\ell \in Y_c \setminus \text{completed} : \ell \text{ has prefix prefix}\}$
        \If{$G \neq \emptyset$} \Comment{Golden match}
            \State $\mathcal{V}_t \gets$ all next tokens after prefix in labels from $G$
        \Else \Comment{Hard negative match}
            \State $\mathcal{V}_t \gets$ next token continuing the hard negative
        \EndIf
        \If{prefix $\in \mathcal{L}_c$} \Comment{Prefix forms a complete label}
            \State $\mathcal{V}_t \gets \mathcal{V}_t \cup \{\texttt{[SEP]}\}$
        \EndIf
    \EndIf
    \State Compute $\ell_t = -\log \left( \frac{\sum_{v \in \mathcal{V}_t} \exp(z_t^{(v)})}{\sum_{v'} \exp(z_t^{(v')})} \right)$
    \If{$y_t = \texttt{[SEP]}$}
        \State $\text{prefix} \gets []$ \Comment{Reset for next label}
    \Else
        \State $\text{prefix} \gets \text{prefix} \oplus [y_t]$ \Comment{Append token}
    \EndIf
    \If{$\text{prefix} \in Y_c$ and $\text{prefix} \notin \text{completed}$}
        \State $\text{completed} \gets \text{completed} \cup \{\text{prefix}\}$
    \EndIf
\EndFor
\State \Return $\frac{1}{T}\sum_{t=1}^T \ell_t$
\end{algorithmic}
\end{algorithm}

\textbf{Training dynamics and efficiency.} Completion losses vanish once the model learns the vocabulary, so loss concentrates at classification positions, where the model cannot rely on prefix correlations and must classify, again if necessary, from the user's history alone, precisely what suppresses exposure bias. 
As the
negatives are precomputed once offline, training stays as cheap as standard teacher forcing, with no generation in the loop. This single mining pass also yields two assets we reuse for the propensity model (\S\ref{sec:dual-encoder}): (1) the per-example negatives, which serve directly as informative negatives for its training, and (2) the hidden states at classification positions, which are \emph{propensity-aligned} by virtue of training, they encode the user's standing interest independent of the generated prefix.

\subsection{Adaptive Patching for Compression}
\label{sec:patching}

Tokenized histories routinely exceed the context budget, forcing a choice between truncating history and losing signal. We adapt the patching strategy of PatchRec~\cite{cui2024multigrained} into a \emph{selective}, recency-aware scheme: rather than compressing the sequence uniformly, we pool token embeddings only within older, less-predictive interaction windows (non-overlapping, window/stride $=10$, $10\times$ compression) while leaving recent interactions at full resolution. This fits a full year of history within a budget that would otherwise hold only the most recent months, preserving fine-grained signal where it matters most. We introduce patching through a four-phase curriculum (Appendix~\ref{app:patching}) that anneals from full to fully-patched sequences, letting the model master the prediction task before adapting to compressed inputs. While the accuracy gain is modest (Table~\ref{tab:patching_ablation}, appendix), the contribution is making extended histories deployable
within fixed inference time token budget limits.

\subsection{Propensity Model}
\label{sec:dual-encoder}

Generator ranks categories at user level, but yields no scores comparable \emph{across} users, so it cannot rank users for a given interest (needed for segmentation and merchandising usecases). We therefore train a lightweight propensity model that outputs cross-user-comparable scores, enabling bidirectional retrieval (interests for users, users for interests) by leveraging the propensity-aligned hidden states and hard negatives surfaced by the exposure-mitigated generator (\S\ref{sec:self-correction}).

\textbf{Architecture.} Self-correction concentrates supervision at classification positions (\S\ref{sec:self-correction}), so the first-token hidden state of each generated category is propensity-aligned and largely prefix-independent. We pass these states for the first $k_g$ generated categories through a small trainable transformer ($\approx$230M params, significantly smaller than the 3.8B generator, which is not trained at this stage) encoder to form a user representation; to score a (user, category) pair, we concatenate this representation with the category's pretrained Cohere~\cite{CohereEmbed2023} embedding and pass it through a shared MLP trained with binary cross-entropy. Concatenation rather than a bare dot product lets the two representations interact through the network, avoiding the single-vector bottleneck we set out to escape. User and category embeddings are each precomputed once and cached, so scoring any pair is a cheap MLP forward pass.

\textbf{Smart negatives.} A generator's confident-but-incorrect predictions are informative hard negatives for free, and we exploit this at \emph{both} pipeline stages. Step~1 negatives from the initial vanilla generator train the exposure-mitigation objective; to supervise the propensity model, we then mine from the \emph{trained} mitigated generator, whose greater diversity surfaces fresh plausible-but-incorrect labels and thus harder negatives than the vanilla model yields. We train the scorer on these alongside popular and random negatives.

\begin{table*}[!t]
\centering
\caption{Generative classification performance (\textbf{P}recision, \textbf{R}ecall, \textbf{F1}) and training-objective
ablation. Absolute improvement over the trivial baseline across product
verticals; The lower block ablates different training-objectives.
}
\label{tab:main_results}
\small
\setlength{\tabcolsep}{3.5pt}
\resizebox{\textwidth}{!}{
\begin{tabular}{@{}l@{\hspace{4pt}}c@{\hspace{4pt}}c@{\hspace{4pt}}c@{\hspace{4pt}}c@{\hspace{4pt}}c@{\hspace{4pt}}c@{\hspace{4pt}}c@{\hspace{4pt}}c@{\hspace{4pt}}c@{\hspace{4pt}}c@{\hspace{4pt}}c@{\hspace{4pt}}c@{\hspace{4pt}}c@{\hspace{4pt}}c@{\hspace{4pt}}c@{\hspace{4pt}}c@{\hspace{4pt}}c@{\hspace{4pt}}c@{}}
\toprule
\multirow{2}{*}{\textbf{Method}} & \multicolumn{3}{c}{\textbf{Global}} & \multicolumn{3}{c}{\textbf{Beauty}} & \multicolumn{3}{c}{\textbf{Apparel}} & \multicolumn{3}{c}{\textbf{Electronics}} & \multicolumn{3}{c}{\textbf{Sports}} & \multicolumn{3}{c}{\textbf{Toys}} \\
\cmidrule(lr){2-4} \cmidrule(lr){5-7} \cmidrule(lr){8-10} \cmidrule(lr){11-13} \cmidrule(lr){14-16} \cmidrule(lr){17-19}
& R & F1 & P & R & F1 & P & R & F1 & P & R & F1 & P & R & F1 & P & R & F1 & P \\
\midrule
\multicolumn{19}{@{}l}{\textit{Baselines}} \\
XLGen & +2.1 & +2.0 & +1.9 & +2.8 & +2.2 & +1.4 & +3.5 & +4.0 & +4.7 & +1.1 & +0.9 & +0.5 & +1.7 & +1.6 & +1.4 & +0.5 & +0.4 & +0.3 \\
GROOV & +1.8 & +2.3 & +2.9 & +2.3 & +2.4 & +2.4 & +3.9 & \textbf{+5.3} & \textbf{+7.5} & +0.2 & +0.3 & +0.6 & +2.2 & +2.6 & \textbf{+3.3} & +2.0 & \textbf{+2.3} & \textbf{+2.7} \\
PatchRec & +0.2 & +1.2 & +3.1 & -2.3 & -2.3 & -2.1 & -0.2 & +1.3 & +5.1 & -1.0 & -0.9 & -0.6 & -1.1 & -0.6 & +0.7 & -1.2 & -1.1 & -0.7 \\
DEXML-DS ($\text{top\_k}{=}5$) & +1.2 & +2.2 & \textbf{+3.8} & +2.6 & +3.1 & \textbf{+3.7} & +1.8 & +2.6 & +3.9 & +1.6 & +2.6 & \textbf{+4.5} & -0.7 & -0.5 & -0.1 & -0.3 & -0.3 & -0.2 \\
DEXML-STk ($\text{top\_k}{=}5$) & -0.9 & -4.1 & -5.2 & +4.1 & -3.1 & -5.7 & -1.8 & -4.2 & -5.1 & -3.2 & -4.4 & -5.4 & -0.9 & -4.4 & -6.0 & -1.5 & -2.2 & -2.5 \\
\midrule
\multicolumn{19}{@{}l}{\textit{Ablation: training objective (ours)}} \\
Standard Teacher Forcing & +0.2 & +0.5 & \textbf{+1.0} & -0.6 & -0.9 & -1.4 & -0.0 & +0.8 & +2.5 & -1.0 & -1.0 & -1.0 & +1.3 & +1.7 & +2.4 & +0.3 & +0.4 & +0.6 \\
Permutation invariance & +3.0 & +1.6 & +0.4 & +4.3 & +1.2 & -1.3 & -0.7 & -0.9 & -1.0 & +0.5 & -0.6 & -1.8 & +2.0 & +0.6 & -0.8 & -0.2 & -0.6 & -1.0 \\
Cross-entropy invariance & +3.4 & +1.9 & +0.6 & +3.9 & +0.6 & -2.0 & -0.4 & -0.7 & -1.2 & +1.5 & +0.3 & -1.1 & +1.8 & +0.4 & -1.1 & -0.0 & -0.4 & -0.9 \\
\midrule
\textbf{MERIT} & \textbf{+14.1} & \textbf{+3.4} & -0.2 & \textbf{+15.5} & \textbf{+3.2} & -1.4 & \textbf{+13.8} & +4.5 & +0.9 & \textbf{+11.5} & \textbf{+2.8} & -0.7 & \textbf{+16.2} & \textbf{+5.1} & +0.5 & \textbf{+7.3} & +1.9 & +0.1 \\
\bottomrule
\end{tabular}
}
\end{table*}

\begin{table*}[!h]
\centering
\caption{Propensity ranking performance (Hit@k). Results show absolute improvement over baseline: DEXML-Stk}
\label{tab:ranking}
\small
\setlength{\tabcolsep}{4.5pt}
\resizebox{\textwidth}{!}{
\begin{tabular}{@{}l@{\hspace{6pt}}c@{\hspace{4pt}}c@{\hspace{4pt}}c@{\hspace{6pt}}c@{\hspace{4pt}}c@{\hspace{4pt}}c@{\hspace{6pt}}c@{\hspace{4pt}}c@{\hspace{4pt}}c@{\hspace{6pt}}c@{\hspace{4pt}}c@{\hspace{4pt}}c@{\hspace{6pt}}c@{\hspace{4pt}}c@{\hspace{4pt}}c@{}}
\toprule
\multirow{2}{*}{\textbf{Method}} & \multicolumn{3}{c}{\textbf{Beauty}} & \multicolumn{3}{c}{\textbf{Apparel}} & \multicolumn{3}{c}{\textbf{Electronics}} & \multicolumn{3}{c}{\textbf{Sports}} & \multicolumn{3}{c}{\textbf{Toys}} \\
\cmidrule(lr){2-4} \cmidrule(lr){5-7} \cmidrule(lr){8-10} \cmidrule(lr){11-13} \cmidrule(lr){14-16}
& @500 & @5k & @20k & @500 & @5k & @20k & @500 & @5k & @20k & @500 & @5k & @20k & @500 & @5k & @20k \\
\midrule
DEXML-DS & +7.8 & +6.8 & +5.2 & +0.4 & +4.4 & +7.5 & -2.4 & +2.9 & +4.7 & -0.4 & -0.9 & +1.1 & -1.6 & -3.3 & -3.0 \\
\textbf{MERIT} (230M) & \textbf{+20.8} & \textbf{+12.6} & \textbf{+21.3} & \textbf{+1.4} & \textbf{+9.3} & \textbf{+19.7} & \textbf{+3.4} & \textbf{+7.1} & \textbf{+14.0} & \textbf{+3.0} & \textbf{+7.8} & \textbf{+15.8} & \textbf{+4.2} & \textbf{+7.4} & \textbf{+15.8} \\
\bottomrule
\end{tabular}
}
\end{table*}

\section{Experiments}
\label{sec:experiments}
We evaluate MERIT on a proprietary e-commerce dataset of anonymized user
interactions (500k training users, 100k disjoint test users), where each
product is annotated with multiple interest categories from a vocabulary
of 291k entries spanning concrete descriptors (e.g.,
``Exercise Equipment'') and intent-level tags (e.g., ``Wellness
Lifestyle''). Inputs and targets follow the formulation of
\S\ref{sec:method}; the year-long history with a 7-day exclusion window
matches production settings where scores are computed ahead of purchase.
We assess three tasks: generative classification (\S\ref{sec:gen-results}),
propensity ranking (\S\ref{sec:prop-results}), and a production A/B test
(\S\ref{sec:abtest}).

\textbf{Setup, Baselines and Metrics.} \label{sec:setup-baselines}
MERIT uses \texttt{phi-3.5-mini} (3.8B) as the base generator, fine-tuned with our exposure-mitigated objective and adaptive patching (refer Appendix~\ref{app:patching}); following~\cite{sun2022clusterguided} we manage the label space via hierarchical generation, organizing the 291k leaf categories into 30k semantic clusters and predicting clusters before leaves. The propensity model is a lightweight 230M-parameter encoder trained on hidden states from the phi generator (\textit{frozen} at this stage), using the first-token hidden states of the first $k_g{=}5$ generated categories per user. Full architecture, hard-negative mining and training details
are deferred to Appendix~\ref{app:impl}.

We compare against representative generative and dual-encoder XMC methods: 
\textbf{Trivial}, predicting user's 5 most frequently purchased categories; \textbf{XLGen}, cluster-guided label generation~\cite{sun2022clusterguided}; \textbf{GROOV}, open-vocabulary generative XMC~\cite{simig2022open}; \textbf{PatchRec}, LLM-based recommendation with patching~\cite{cui2024multigrained}; and a dual encoder (same phi model used) in two loss configurations, \textbf{DEXML-DS} (DecoupledSoftmax) and \textbf{DEXML-STk} (SoftTop-$k$) as proposed by~\cite{dhillon2024dual}. 
Per internal disclosure policy, we report relative gain rather than absolute numbers: For generative classification, we report \textbf{R}ecall/\textbf{P}recision/\textbf{F1} (soft matching via Cohere cosine similarity $\geq0.7$, calibrated on manually judged boundary pairs) absolute improvement over the Trivial baseline (Table~\ref{tab:main_results}); for propensity ranking, we report Hit@$k$ (the fraction of an interest cluster's true-positive users falling in its top-$k$) as improvement over DEXML-STk (Table~\ref{tab:ranking}). We validate this soft-matching protocol directly with human judgments and a threshold-sensitivity sweep in Appendix~\ref{app:threshold}: at $\tau{=}0.7$, human annotators judged 96.2\% of a 230-pair boundary sample as at least loosely semantically equivalent (65.4\% strictly equivalent), and MERIT's recall advantage over every baseline holds under all matching criteria we tested, including exact match.

\subsection{Generative Classification and Ablations}
\label{sec:gen-results}
Table~\ref{tab:main_results} reports generative-classification performance and ablates the training objective; we show the five most important business verticals here, with all 20 in Appendix~\ref{app:full-gen} (Tables~\ref{tab:ext_genA} - \ref{tab:ext_genC}). \textit{Global} pools predictions over the entire test set rather than averaging the shown verticals. MERIT improves Global recall by $+14.1$ over Trivial and by at least $+11.9$ over the strongest baseline (the conservative figure reported in the abstract), with consistent F1 gains. 
Dual encoders rank the full label space and need a top-$k$ cutoff to yield predictions. We tuned for F1 on a control set, this gives $k{=}5$ (the value at which DEXML baselines are reported), and even so they perform poorly. Raising $k$ to 100 still trails MERIT's recall by $+1.8$ despite severely low precision, pointing to embedding collapse: an overloaded space favoring one dominant category and its neighbors, the single-vector analogue of the correlation cascade, rather than output budget.
Precision dips slightly in some verticals, most notably Apparel, where GROOV edges our F1/precision despite our $3.5\times$ higher recall. This is an acceptable trade-off since recall is the operative retrieval metric: missed interests are unrecoverable downstream, while over-generation is cheaply filtered. This is also supported by
\S\ref{sec:abtest}. 

The ablation isolates the gain's source: \textit{Permutation invariance} (similar to~\cite{simig2022open}) and \textit{Cross-entropy Invariance} (the same, with cross-entropy loss instead) each add only modest recall over Standard Teacher Forcing compared to our objective's $+13.9$ recall (same model/data/architecture). To further reinforce the objective's benefit, MERIT still outperforms aggregating four independent temperature-sampled outputs from the Standard Teacher Forcing model, in both recall ($+6.75$) and precision ($+1.76$), 
ruling out the alternate explanation that gains merely stem from diversity but rather from recovery from erroneous prefixes. 

\subsection{Propensity Model based Ranking}
\label{sec:prop-results}
Ranking users for individual fine-grained categories is statistically unstable:
most categories have very few true-positive users. We therefore evaluate
at a much coarser product-category granularity, also the decision-relevant
level for campaign targeting. We aggregate leaf propensities as
$P(g \mid u) = \sum_{c} P(g \mid c) \, P(c \mid u)$, where $P(g\mid c)$ is the empirical likelihood
that category $c$ implies
product-category $g$ (precomputed offline) and $P(c\mid u)$ comes from
the propensity model. Table~\ref{tab:ranking} compares against the same
3.8B \texttt{phi-3.5-mini} dual encoder~\cite{dhillon2024dual}. Full per-vertical results across all 20 verticals and every Hit@$k$ threshold are in Appendix~\ref{app:full-rank} (Table~\ref{tab:ext_ranking}). 


Averaged over all $k$ and verticals, MERIT's 230M encoder trained on \emph{frozen} generator embeddings improves Hit@$k$ by $+6.1$ (figure reported in abstract) over DEXML-DS despite far fewer parameters, which we attribute to its less-collapsed, propensity-aligned embeddings and the mined negatives from the generator. Crucially, its scores are comparable across users, enabling both user$\rightarrow$category (Table~\ref{tab:main_results}) and category$\rightarrow$user retrieval (Table~\ref{tab:ranking}), showing our exposure mitigated generator unlocks bidirectional ranking. Appendix~\ref{app:kg} reports a full sensitivity sweep over $k_g$ (the number of generated categories pooled into the user representation) and we see that the ranking gains above are not sensitive to this choice.

\subsection{Production A/B Test}
\label{sec:abtest}

We tested MERIT in online A/B tests selecting users for targeted
merchandising campaigns from their predicted propensities. Across 3
one-week campaigns in different verticals (0.6M - 3.5M customers per
treatment arm), users were randomly assigned to \textbf{T}reatment or
\textbf{C}ontrol, with T segments based on user - campaign propensity and
C segments configured by domain experts. Treatments were matched in
size, creative, and delivery.
Lift was evaluated using a Bayesian, covariate-adjusted online-experiment framework~\cite{kamalbasha2020bayesianabtestingbusiness, 10.1145/2740908.2742563}, with the posterior probability of positive lift,
$P(\mathrm{lift}>0)$, as the decision metric.
All campaigns cleared our internal shipping
threshold of $(P(\mathrm{lift}>0)\ge 0.90)$ with per-campaign lift of:
Personal Care $+0.23\%$, Toys $+0.19\%$, and Sports $+0.37\%$
($+0.26\%$ avg.).
Detailed credible intervals 
are reported in App.~\ref{app:abtest}
(Table~\ref{tab:abtest}).
Deployment is two-stage: user embeddings are extracted periodically and stored, and interest$\rightarrow$user scoring 
reduces to lightweight MLP lookup over precomputed category embeddings, making the system well-suited to our usecase.

\section{Conclusion}

We presented MERIT, an efficient exposure-mitigated training framework for generative extreme multi-label classification in a user-interest retrieval setting. A single-pass self-correction objective teaches generators to recover from erroneous prefixes without the per-step cost of RL or scheduled sampling, yielding propensity-aligned embeddings
that power a lightweight scorer for offline customer segmentation ahead of time. It outperforms strong generative and dual-encoder baselines and delivers meaningful lift in production settings. 
Since exposure bias may arise in any teacher-forced generator operating over large, structured label spaces, 
the proposed method
offers a general mechanism for improving robustness in generative XMC and multi-label generation tasks.

\section{Limitations}

MERIT trades a portion of precision to optimize for recall gains across all verticals, by design: recall is the operative metric for an upstream retrieval interface, but the raw generated set is noisier and may require downstream filtering before use in precision-sensitive applications. 

Our evaluation is on a single proprietary e-commerce dataset, and per internal disclosure policy we report relative rather than absolute metrics, limiting direct comparison to public XMC benchmarks and other generative-XMC literature. Extending MERIT to public XMC benchmarks is left to future work. 

Because our target (gold labels) reflect only categories corresponding to a user's immediate next purchase, mined negatives from \S\ref{sec:self-correction} can occasionally be unobserved true interests; training and evaluating over multiple future purchases rather than a single next-purchase target could yield more stable propensities, and is left to future work


Finally, as per the current business requirement, our deployment is two-stage and offline: user embeddings are extracted and scored in batch, so the system supports segmentation and merchandising use cases. Real-time, on-demand scoring while has its benefits has substantial cost implications as well, and a thorough analysis is left to future work.

\section*{Ethical Considerations}

MERIT is trained on anonymized user-interaction logs, and interest-category descriptors are derived from product metadata rather than from personal or demographic attributes; we do not model or predict sensitive attributes (e.g., health, financial, or demographic status) as interest categories. Predicted propensities are not shown directly to customers: they feed a downstream business-rule and eligibility-filtering layer (\S\ref{sec:intro}) before influencing any customer-facing surface, which also bounds the impact of MERIT's recall-oriented, higher-noise predictions.

\section*{Acknowledgments}
We thank the anonymous reviewers and area chair for their constructive feedback, which improved this paper.

\bibliography{custom}

\appendix

\clearpage
\appendix
\twocolumn[
  \begin{center}
    {\Large\bfseries Appendix}
  \end{center}
  \vspace{1em}
]

\section{Detailed Walkthrough of Exposure Mitigation Training}
\label{app:dry-run}

To illustrate the exposure mitigation training procedure (Algorithm~\ref{alg:self-correct}), we present a complete example showing how the self-correction objective operates at each token position. This walkthrough uses all variables from the algorithm to clarify their roles.

\subsection{Setup}

\textbf{User Context:} Consider a user with profile frequent sports purchases, occasional home items. Let's say, they purchased a Cricket Bat. Therefore,

\textbf{Golden Labels} (ground truth interest categories for the purchased product):

\begin{equation*}
\begin{aligned}
Y_c = \{\text{"Sports"}, \text{"Sports Enthusiast"}, \\
\text{"Cricket Fan"}, \text{"Sports Gear"}\}
\end{aligned}
\end{equation*}

\textbf{Mined Hard Negatives} (incorrect but plausible predictions from stage-1 model):
\begin{equation*}
\hat{Y}_{\text{filtered}} = \{\text{"Home Decor"}, \text{"Milk"}, \text{"Mobile Phone"}\}
\end{equation*}

\textbf{Augmented Label Set:}
\begin{equation*}
\mathcal{L}_c = Y_c \cup \hat{Y}_{\text{filtered}}
\end{equation*}

\textbf{Tokenization} (simplified for clarity):
\begin{itemize}[leftmargin=*,itemsep=2pt]
    \item "Sports" $\rightarrow$ \texttt{[Sports]}
    \item "Sports Enthusiast" $\rightarrow$ \texttt{[Sports, Enthusiast]}
    \item "Cricket Fan" $\rightarrow$ \texttt{[Cricket, Fan]}
    \item "Sports Gear" $\rightarrow$ \texttt{[Sports, Gear]}
    \item "Home Decor" $\rightarrow$ \texttt{[Home, Decor]}
    \item "Milk" $\rightarrow$ \texttt{[Milk]}
    \item "Mobile Phone" $\rightarrow$ \texttt{[Mobile, Phone]}
\end{itemize}

\textbf{Shuffled Target Sequence} (after random shuffle of $\mathcal{L}_c$):
\begin{equation*}
\begin{aligned}
S = [&\text{Home}, \text{Decor}, \texttt{[SEP]}, \text{Sports}, \texttt{[SEP]}, \text{Milk}, \texttt{[SEP]}, \\ 
&\text{Cricket}, \text{Fan}, \texttt{[SEP]}, \text{Sports}, \text{Enthusiast}, \texttt{[SEP]}, \\
&\text{Mobile}, \text{Phone}, \texttt{[SEP]}, \text{Sports}, \text{Gear}, \texttt{[END]}]
\end{aligned}
\end{equation*}

\subsection{Forward Pass with Loss Computation}

\textbf{Important:} We perform a \emph{single forward pass} through the model with the input prompt concatenated with the shuffled target sequence $S$. This produces next-token logits $z_1, z_2, \ldots, z_{19}$ for each position. We then compute the loss at each position according to our exposure mitigation rules. The model does \emph{not} autoregressively produce tokens during training, this is close to teacher forcing, just with modified loss computation that decides which target tokens probabilities should be penalized.

\textbf{Input to model:} 
\begin{equation*}
\text{[User Profile] [Recent Interactions] } S
\end{equation*}

\textbf{Model output:} Logits $z_1, z_2, \ldots, z_{19}$ representing next-token probability distributions at each position.

We now walk through each position $t$ in the target sequence $S$, showing how we compute the loss $\ell_t$ on the corresponding logits $z_t$.

\vspace{1em}
\noindent\textbf{Position $t=1$: Token $y_1 = $ "Home", Logits $z_1$}

\textbf{State:}
\begin{itemize}[leftmargin=*,itemsep=1pt]
    \item $\text{prefix} = []$ (empty, starting new label)
    \item $\text{completed} = \{\}$ (no golden labels completed yet)
    \item $Y_c \setminus \text{completed} = \{ \begin{aligned}[t] 
    &\text{Sports, Sports Enthusiast,} \\ 
    &\text{Cricket Fan, Sports Gear} \} \end{aligned}$
\end{itemize}

\textbf{Valid targets $\mathcal{V}_1$:} Since prefix is empty, we are at a \emph{classification position}. Valid targets are first tokens of all uncompleted golden labels:
\begin{equation*}
\mathcal{V}_1 = \{\text{Sports}, \text{Cricket}\}
\end{equation*}
Note: "Sports" appears as the first token of three labels (Sports, Sports Enthusiast, Sports Gear), and "Cricket" is the first token of Cricket Fan.

\textbf{Loss computation on $z_1$:}
\begin{equation*}
\ell_1 = -\log \left( \frac{\exp(z_1^{\text{Sports}}) + \exp(z_1^{\text{Cricket}})}{\sum_{v'} \exp(z_1^{(v')})} \right)
\end{equation*}

\textbf{Discussion:} This is a classification position where we supervise the model to predict correct interest categories based on user context. Although the actual next token in the sequence is "Home" (incorrect), we only reward the logits corresponding to "Sports" and "Cricket" (golden first tokens). Note, the loss is permutation invariant in the sense that we are not penalizing the model if it is giving high probability to either Sports or Cricket.

\textbf{State update:} prefix $\gets$ [Home]

\vspace{1em}
\noindent\textbf{Position $t=2$: Token $y_2 = $ "Decor", Logits $z_2$}

\textbf{State:}
\begin{itemize}[leftmargin=*,itemsep=1pt]
    \item $\text{prefix} = [\text{Home}]$
    \item $\text{completed} = \{\}$
    \item $Y_c \setminus \text{completed} = \{ \begin{aligned}[t] 
    &\text{Sports, Sports Enthusiast,} \\ 
    &\text{Cricket Fan, Sports Gear} \} \end{aligned}$
\end{itemize}

\textbf{Valid targets $\mathcal{V}_2$:} Prefix is non-empty. Check if it matches any golden labels:
\begin{equation*}
G = \{\ell \in Y_c \setminus \text{completed} : \ell \text{ starts with "Home"}\} = \emptyset
\end{equation*}
No golden labels start with "Home". This prefix matches the hard negative "Home Decor". For \emph{label coherence}, we complete it:
\begin{equation*}
\mathcal{V}_2 = \{\text{Decor}\}
\end{equation*}

\textbf{Loss computation on $z_2$:}
\begin{equation*}
\ell_2 = -\log \left( \frac{\exp(z_2^{\text{Decor}})}{\sum_{v'} \exp(z_2^{(v')})} \right)
\end{equation*}

\textbf{Discussion:} Once the prefix becomes "Home" (incorrect), we enforce completion of the full phrase "Home Decor" rather than allowing the loss to reward switching mid-label. This prevents malformed outputs and teaches the model to produce well-formed categories even when the prefix corresponds to an error.

\textbf{State update:} prefix $\gets$ [Home, Decor]

\vspace{1em}
\noindent\textbf{Position $t=3$: Token $y_3 = \texttt{[SEP]}$, Logits $z_3$}

\textbf{State:}
\begin{itemize}[leftmargin=*,itemsep=1pt]
    \item $\text{prefix} = [\text{Home, Decor}]$
    \item Prefix forms complete label "Home Decor" $\in \mathcal{L}_c$
    \item $\text{completed} = \{\}$ ("Home Decor" is a hard negative, not golden)
\end{itemize}

\textbf{Valid targets $\mathcal{V}_3$:} The prefix completes a label (the hard negative), so separator is valid:
\begin{equation*}
\mathcal{V}_3 = \{\texttt{[SEP]}\}
\end{equation*}

\textbf{Loss computation on $z_3$:}
\begin{equation*}
\ell_3 = -\log \left( \frac{\exp(z_3^{\texttt{[SEP]}})}{\sum_{v'} \exp(z_3^{(v')})} \right)
\end{equation*}

\textbf{State update:} Since $y_3 = \texttt{[SEP]}$, reset prefix $\gets []$

\vspace{1em}
\noindent\textbf{Position $t=4$: Token $y_4 = $ "Sports", Logits $z_4$}

\textbf{State:}
\begin{itemize}[leftmargin=*,itemsep=1pt]
    \item $\text{prefix} = []$ (reset after separator)
    \item $\text{completed} = \{\}$
    \item $Y_c \setminus \text{completed} = \{ \begin{aligned}[t] 
    &\text{Sports, Sports Enthusiast,} \\ 
    &\text{Cricket Fan, Sports Gear} \} \end{aligned}$
\end{itemize}

\textbf{Valid targets $\mathcal{V}_4$:} Prefix empty, so this is a \emph{classification position}:
\begin{equation*}
\mathcal{V}_4 = \{\text{Sports}, \text{Cricket}\}
\end{equation*}

\textbf{Loss computation on $z_4$:}
\begin{equation*}
\ell_4 = -\log \left( \frac{\exp(z_4^{\text{Sports}}) + \exp(z_4^{\text{Cricket}})}{\sum_{v'} \exp(z_4^{(v')})} \right)
\end{equation*}

\textbf{Discussion:} This is another classification position, but now the consumed prefix contains "Home Decor" (incorrect). The model must learn to make independent classification decisions despite having seen errors in the prefix. This is the core of \emph{exposure mitigation}: training the model to recover from mistakes that appear in the prefix.

\textbf{State update:} prefix $\gets$ [Sports]

\vspace{1em}
\noindent\textbf{Position $t=5$: Token $y_5 = \texttt{[SEP]}$, Logits $z_5$}

\textbf{State:}
\begin{itemize}[leftmargin=*,itemsep=1pt]
    \item $\text{prefix} = [\text{Sports}]$
    \item Prefix "Sports" is a complete golden label AND partial match for others
    \item $\text{completed} = \{\}$
\end{itemize}

\textbf{Valid targets $\mathcal{V}_5$:} Prefix "Sports" matches multiple golden labels:
\begin{equation*}
G = \{\text{Sports, Sports Enthusiast, Sports Gear}\}
\end{equation*}
Valid continuations include tokens that extend any of these labels:
\begin{itemize}[leftmargin=*,itemsep=1pt]
    \item "Sports" is already complete $\rightarrow$ \texttt{[SEP]} is valid
    \item "Sports Enthusiast" needs "Enthusiast"
    \item "Sports Gear" needs "Gear"
\end{itemize}
Therefore:
\begin{equation*}
\mathcal{V}_5 = \{\texttt{[SEP]}, \text{Enthusiast}, \text{Gear}\}
\end{equation*}

\begin{multline*}
\ell_5 = -\log \Bigg(
  \big(\exp(z_5^{\texttt{[SEP]}}) + \exp(z_5^{\text{Enthusiast}}) \\
  + \exp(z_5^{\text{Gear}})\big) \Big/ {\textstyle\sum_{v'} \exp(z_5^{(v')})}
\Bigg)
\end{multline*}

\textbf{Discussion:} This position demonstrates \emph{permutation invariance} and robustness to edge cases. We pool probability mass over all valid options: treating "Sports" as complete, or extending it to "Sports Enthusiast" or "Sports Gear". This avoids penalizing any particular ordering choice.

\textbf{State update:} Since $y_5 = \texttt{[SEP]}$, prefix $\gets []$ and completed $\gets \{\text{Sports}\}$

\vspace{1em}
\noindent\textbf{Position $t=6$: Token $y_6 = $ "Milk", Logits $z_6$}

\textbf{State:}
\begin{itemize}[leftmargin=*,itemsep=1pt]
    \item $\text{prefix} = []$
    \item $\text{completed} = \{\text{Sports}\}$
    \item $Y_c \setminus \text{completed} = \{ \begin{aligned}
    [t]
    &\text{Sports Enthusiast,}  \\
    &\text{Cricket Fan, Sports Gear}\} \end{aligned}$
    
\end{itemize}

\textbf{Valid targets $\mathcal{V}_6$:} Classification position, first tokens of remaining golden labels:
\begin{equation*}
\mathcal{V}_6 = \{\text{Sports}, \text{Cricket}\}
\end{equation*}

\textbf{Loss computation on $z_6$:}
\begin{equation*}
\ell_6 = -\log \left( \frac{\exp(z_6^{\text{Sports}}) + \exp(z_6^{\text{Cricket}})}{\sum_{v'} \exp(z_6^{(v')})} \right)
\end{equation*}

\textbf{Discussion:} Another classification position after the prefix has consumed both correct ("Sports") and incorrect ("Home Decor") labels. The loss trains the model to classify independently of prefix correlations.

\textbf{State update:} prefix $\gets$ [Milk]

\vspace{1em}
\noindent\textbf{Position $t=7$: Token $y_7 = \texttt{[SEP]}$, Logits $z_7$}

\textbf{State:}
\begin{itemize}[leftmargin=*,itemsep=1pt]
    \item $\text{prefix} = [\text{Milk}]$
    \item "Milk" is a complete hard negative label
\end{itemize}

\textbf{Valid targets $\mathcal{V}_7$:}
\begin{equation*}
\mathcal{V}_7 = \{\texttt{[SEP]}\}
\end{equation*}

\textbf{Loss computation on $z_7$:}
\begin{equation*}
\ell_7 = -\log \left( \frac{\exp(z_7^{\texttt{[SEP]}})}{\sum_{v'} \exp(z_7^{(v')})} \right)
\end{equation*}

\textbf{State update:} prefix $\gets []$

\vspace{1em}
\noindent\textbf{Position $t=8$: Token $y_8 = $ "Cricket", Logits $z_8$}

\textbf{State:}
\begin{itemize}[leftmargin=*,itemsep=1pt]
    \item $\text{prefix} = []$
    \item $\text{completed} = \{\text{Sports}\}$
    \item $Y_c \setminus \text{completed} = \{ \begin{aligned}
    [t]
    &\text{Sports Enthusiast,}  \\
    &\text{Cricket Fan, Sports Gear}\} \end{aligned}$
\end{itemize}

\textbf{Valid targets $\mathcal{V}_8$:}
\begin{equation*}
\mathcal{V}_8 = \{\text{Sports}, \text{Cricket}\}
\end{equation*}

\textbf{Loss computation on $z_8$:}
\begin{equation*}
\ell_8 = -\log \left( \frac{\exp(z_8^{\text{Sports}}) + \exp(z_8^{\text{Cricket}})}{\sum_{v'} \exp(z_8^{(v')})} \right)
\end{equation*}

\textbf{State update:} prefix $\gets$ [Cricket]

\vspace{1em}
\noindent\textbf{Position $t=9$: Token $y_9 = $ "Fan", Logits $z_9$}

\textbf{State:}
\begin{itemize}[leftmargin=*,itemsep=1pt]
    \item $\text{prefix} = [\text{Cricket}]$
    \item $\text{completed} = \{\text{Sports}\}$
\end{itemize}

\textbf{Valid targets $\mathcal{V}_9$:} Prefix "Cricket" matches exactly one golden label "Cricket Fan":
\begin{equation*}
G = \{\text{Cricket Fan}\} \quad \Rightarrow \quad \mathcal{V}_9 = \{\text{Fan}\}
\end{equation*}

\textbf{Loss computation on $z_9$:}
\begin{equation*}
\ell_9 = -\log \left( \frac{\exp(z_9^{\text{Fan}})}{\sum_{v'} \exp(z_9^{(v')})} \right)
\end{equation*}

\textbf{State update:} prefix $\gets$ [Cricket, Fan]

\vspace{1em}
\noindent\textbf{Position $t=10$: Token $y_{10} = \texttt{[SEP]}$, Logits $z_{10}$}

\textbf{State:}
\begin{itemize}[leftmargin=*,itemsep=1pt]
    \item $\text{prefix} = [\text{Cricket, Fan}]$
    \item Prefix completes golden label "Cricket Fan"
\end{itemize}

\textbf{Valid targets $\mathcal{V}_{10}$:}
\begin{equation*}
\mathcal{V}_{10} = \{\texttt{[SEP]}\}
\end{equation*}

\textbf{Loss computation on $z_{10}$:}
\begin{equation*}
\ell_{10} = -\log \left( \frac{\exp(z_{10}^{\texttt{[SEP]}})}{\sum_{v'} \exp(z_{10}^{(v')})} \right)
\end{equation*}

\textbf{State update:} prefix $\gets []$, completed $\gets \{\text{Sports, Cricket Fan}\}$

\vspace{1em}
\noindent\textbf{Position $t=11$: Token $y_{11} = $ "Sports", Logits $z_{11}$}

\textbf{State:}
\begin{itemize}[leftmargin=*,itemsep=1pt]
    \item $\text{prefix} = []$
    \item $\text{completed} = \{\text{Sports, Cricket Fan}\}$
    \item $Y_c \setminus \text{completed} = \{ \begin{aligned}
    [t]
    &\text{Sports Enthusiast,}  \\
    &\text{Sports Gear}\} \end{aligned}$
\end{itemize}

\textbf{Valid targets $\mathcal{V}_{11}$:} Both remaining golden labels start with "Sports":
\begin{equation*}
\mathcal{V}_{11} = \{\text{Sports}\}
\end{equation*}

\textbf{Loss computation on $z_{11}$:}
\begin{equation*}
\ell_{11} = -\log \left( \frac{\exp(z_{11}^{\text{Sports}})}{\sum_{v'} \exp(z_{11}^{(v')})} \right)
\end{equation*}

\textbf{State update:} prefix $\gets$ [Sports]

\vspace{1em}
\noindent\textbf{Position $t=12$: Token $y_{12} = $ "Enthusiast", Logits $z_{12}$}

\textbf{State:}
\begin{itemize}[leftmargin=*,itemsep=1pt]
    \item $\text{prefix} = [\text{Sports}]$
    \item $\text{completed} = \{\text{Sports, Cricket Fan}\}$
    \item Remaining: $\{\text{Sports Enthusiast, Sports Gear}\}$
\end{itemize}

\textbf{Valid targets $\mathcal{V}_{12}$:} Prefix "Sports" matches both remaining labels:
\begin{equation*}
\begin{aligned}
 G = \{\text{Sports Enthusiast, Sports Gear}\} \\
 \quad \Rightarrow \quad \mathcal{V}_{12} = \{\text{Enthusiast}, \text{Gear}\}   
\end{aligned}
\end{equation*}

\textbf{Loss computation on $z_{12}$:}
\begin{equation*}
\ell_{12} = -\log \left( \frac{\exp(z_{12}^{\text{Enthusiast}}) + \exp(z_{12}^{\text{Gear}})}{\sum_{v'} \exp(z_{12}^{(v')})} \right)
\end{equation*}

\textbf{Discussion:} Permutation invariance, the loss pools probability over both "Enthusiast" and "Gear" since either continuation is valid.

\textbf{State update:} prefix $\gets$ [Sports, Enthusiast]

\vspace{1em}
\noindent\textbf{Position $t=13$: Token $y_{13} = \texttt{[SEP]}$, Logits $z_{13}$}

\textbf{State:}
\begin{itemize}[leftmargin=*,itemsep=1pt]
    \item $\text{prefix} = [\text{Sports, Enthusiast}]$
    \item Prefix completes "Sports Enthusiast"
\end{itemize}

\textbf{Valid targets $\mathcal{V}_{13}$:}
\begin{equation*}
\mathcal{V}_{13} = \{\texttt{[SEP]}\}
\end{equation*}

\textbf{Loss computation on $z_{13}$:}
\begin{equation*}
\ell_{13} = -\log \left( \frac{\exp(z_{13}^{\texttt{[SEP]}})}{\sum_{v'} \exp(z_{13}^{(v')})} \right)
\end{equation*}

\textbf{State update:} prefix $\gets []$, \\ completed $\gets \{\text{Sports, Cricket Fan, Sports Enthusiast}\}$

\vspace{1em}
\noindent\textbf{Position $t=14$: Token $y_{14} = $ "Mobile", Logits $z_{14}$}

\textbf{State:}
\begin{itemize}[leftmargin=*,itemsep=1pt]
    \item $\text{prefix} = []$ (reset after separator)
    \item $\text{completed} = \{ \begin{aligned} 
    [t]
    &\text{Sports, Cricket Fan,} \\
    &\text{Sports Enthusiast}\} \end{aligned}$
    \item $Y_c \setminus \text{completed} = \{\text{Sports Gear}\}$
\end{itemize}

\textbf{Valid targets $\mathcal{V}_{14}$:} Classification position, only one golden label remains:
\begin{equation*}
\mathcal{V}_{14} = \{\text{Sports}\}
\end{equation*}
Only "Sports Gear" remains, which starts with "Sports".

\textbf{Loss computation on $z_{14}$:}
\begin{equation*}
\ell_{14} = -\log \left( \frac{\exp(z_{14}^{\text{Sports}})}{\sum_{v'} \exp(z_{14}^{(v')})} \right)
\end{equation*}

\textbf{Discussion:} Another classification position demonstrating exposure mitigation. 

\textbf{State update:} prefix $\gets$ [Mobile]

\vspace{1em}
\noindent\textbf{Position $t=15$: Token $y_{15} = $ "Phone", Logits $z_{15}$}

\textbf{State:}
\begin{itemize}[leftmargin=*,itemsep=1pt]
    \item $\text{prefix} = [\text{Mobile}]$
    \item $\text{completed} = \{ \begin{aligned} 
    [t]
    &\text{Sports, Cricket Fan,} \\
    &\text{Sports Enthusiast}\} \end{aligned}$

\end{itemize}

\textbf{Valid targets $\mathcal{V}_{15}$:} Prefix "Mobile" matches hard negative "Mobile Phone":
\begin{equation*}
G = \emptyset \quad \Rightarrow \quad \mathcal{V}_{15} = \{\text{Phone}\}
\end{equation*}
No golden labels match, so complete the hard negative for coherence.

\textbf{Loss computation on $z_{15}$:}
\begin{equation*}
\ell_{15} = -\log \left( \frac{\exp(z_{15}^{\text{Phone}})}{\sum_{v'} \exp(z_{15}^{(v')})} \right)
\end{equation*}

\textbf{State update:} prefix $\gets$ [Mobile, Phone]

\vspace{1em}
\noindent\textbf{Position $t=16$: Token $y_{16} = \texttt{[SEP]}$, Logits $z_{16}$}

\textbf{State:}
\begin{itemize}[leftmargin=*,itemsep=1pt]
    \item $\text{prefix} = [\text{Mobile, Phone}]$
    \item "Mobile Phone" is complete
\end{itemize}

\textbf{Valid targets $\mathcal{V}_{16}$:}
\begin{equation*}
\mathcal{V}_{16} = \{\texttt{[SEP]}\}
\end{equation*}

\textbf{Loss computation on $z_{16}$:}
\begin{equation*}
\ell_{16} = -\log \left( \frac{\exp(z_{16}^{\texttt{[SEP]}})}{\sum_{v'} \exp(z_{16}^{(v')})} \right)
\end{equation*}

\textbf{State update:} prefix $\gets []$

\vspace{1em}
\noindent\textbf{Position $t=17$: Token $y_{17} = $ "Sports", Logits $z_{17}$}

\textbf{State:}
\begin{itemize}[leftmargin=*,itemsep=1pt]
    \item $\text{prefix} = []$
    \item $\text{completed} = \{ \begin{aligned} 
    [t]
    &\text{Sports, Cricket Fan,} \\
    &\text{Sports Enthusiast}\} \end{aligned}$
    \item $Y_c \setminus \text{completed} = \{\text{Sports Gear}\}$
\end{itemize}

\textbf{Valid targets $\mathcal{V}_{17}$:} Only one golden label remains:
\begin{equation*}
\mathcal{V}_{17} = \{\text{Sports}\}
\end{equation*}

\textbf{Loss computation on $z_{17}$:}
\begin{equation*}
\ell_{17} = -\log \left( \frac{\exp(z_{17}^{\text{Sports}})}{\sum_{v'} \exp(z_{17}^{(v')})} \right)
\end{equation*}

\textbf{State update:} prefix $\gets$ [Sports]

\vspace{1em}
\noindent\textbf{Position $t=18$: Token $y_{18} = $ "Gear", Logits $z_{18}$}

\textbf{State:}
\begin{itemize}[leftmargin=*,itemsep=1pt]
    \item $\text{prefix} = [\text{Sports}]$
    \item $\text{completed} = \{ \begin{aligned} 
    [t]
    &\text{Sports, Cricket Fan,} \\
    &\text{Sports Enthusiast}\} \end{aligned}$
    \item Remaining: $\{\text{Sports Gear}\}$
\end{itemize}

\textbf{Valid targets $\mathcal{V}_{18}$:} Prefix "Sports" matches the last remaining golden label:
\begin{equation*}
G = \{\text{Sports Gear}\} \quad \Rightarrow \quad \mathcal{V}_{18} = \{\text{Gear}\}
\end{equation*}

\textbf{Loss computation on $z_{18}$:}
\begin{equation*}
\ell_{18} = -\log \left( \frac{\exp(z_{18}^{\text{Gear}})}{\sum_{v'} \exp(z_{18}^{(v')})} \right)
\end{equation*}

\textbf{State update:} prefix $\gets$ [Sports, Gear]

\vspace{1em}
\noindent\textbf{Position $t=19$: Token $y_{19} = \texttt{[END]}$, Logits $z_{19}$}

\textbf{State:}
\begin{itemize}[leftmargin=*,itemsep=1pt]
    \item $\text{prefix} = [\text{Sports, Gear}]$
    \item $\text{completed} = \{ \begin{aligned} 
    [t]
    &\text{Sports, Cricket Fan,} \\
    &\text{Sports Enthusiast}\} \end{aligned}$
    \item All golden labels will be complete after this position
\end{itemize}

\textbf{Valid targets $\mathcal{V}_{19}$:} "Sports Gear" is complete. Since all golden labels are now completed, only the end token is valid:
\begin{equation*}
\mathcal{V}_{19} = \{\texttt{[END]}\}
\end{equation*}

\textbf{Loss computation on $z_{19}$:}
\begin{equation*}
\ell_{19} = -\log \left( \frac{\exp(z_{19}^{\texttt{[END]}})}{\sum_{v'} \exp(z_{19}^{(v')})} \right)
\end{equation*}

\textbf{State update:} \\ completed $\gets \{ \begin{aligned}
[t] &\text{Sports, Cricket Fan,} \\ &\text{Sports Enthusiast, Sports Gear}\} \end{aligned}$ (all golden labels complete)

\vspace{1em}
\noindent\textbf{Final Loss:}
\begin{equation*}
\mathcal{L} = \frac{1}{19} \sum_{t=1}^{19} \ell_t
\end{equation*}

\subsection{Key Insights}

\textbf{Exposure mitigation:} Classification positions (t=1,4,6,8,11,14,17) occur after the prefix has consumed both correct and incorrect labels. The model learns to make independent classifications based on user context rather than following prefix correlations. Crucially, this means hidden states at classification positions encode genuine user-interest propensities independent of the prefix, explaining why the Propensity Model achieves strong ranking performance despite using representations extracted from a generative task. These representations are propensity-aligned by design.

\textbf{Permutation invariance:} Positions like t=5,12 pool probability mass over multiple valid continuations, avoiding penalties for arbitrary ordering choices.

\textbf{Label coherence:} Positions like t=2,7 enforce completion of hard negatives once started, preventing malformed outputs.

\textbf{Training dynamics:} Early in training, completion losses (t=2,9,12) are high as the model learns the label vocabulary. These quickly decrease. Classification losses (t=1,4,6,8,11) remain challenging throughout training as the model learns which interests match the user profile from 250k+ candidates. This natural concentration of learning signal at classification positions is exactly what suppresses exposure bias.

\section{Threshold Sensitivity and Human Agreement}
\label{app:threshold}

To validate the $\tau{=}0.7$ soft-matching protocol used throughout the paper, we (1) re-evaluate all methods across a range of thresholds and under normalized Exact Match, and (2) collect human judgments on a sample of boundary pairs.

\textbf{Threshold sensitivity.} Table~\ref{tab:threshold_sensitivity} is from an independent re-evaluation run conducted for this analysis; absolute values at $\tau{=}0.70$ therefore differ marginally (within $0.6$pp) from Table~\ref{tab:main_results}'s soft-matching numbers due to ordinary run-to-run variance, with Table~\ref{tab:main_results} the authoritative source for the paper's headline results. Method rankings and conclusions are unchanged. The table reports absolute improvement over the Trivial baseline, pooled globally across verticals, at $\tau \in \{0.70, 0.80, 0.90\}$ and under normalized Exact Match. $\tau$-AUC is the normalized trapezoidal area under the improvement-versus-threshold curve over $\tau \in [0.70, 0.90]$. MERIT's recall advantage holds under every criterion, including Exact Match, and relative rankings across methods are unchanged across $\tau$. Some baselines edge out MERIT on precision at stricter thresholds, which is acceptable for the retrieval setting of this work: recall is our operating metric for the generative model, as downstream filtering is cheap and missed candidates are irrecoverable. Propensity ranking (Table~\ref{tab:ranking}, main paper) and the A/B campaigns are unaffected by this choice, since they aggregate leaf propensities via the structural likelihood in \S\ref{sec:prop-results}, not soft-matching.

\textbf{Human agreement.} We sampled 230 interest-category pairs directly around the $\tau{=}0.7$ decision boundary to evaluate the hardest edge cases, rather than an uninformative random sample. Human annotators evaluating semantic equivalence judged 96.2\% of pairs accepted at $\tau{=}0.7$ as at least loosely equivalent (similar purchase intent), with 65.4\% judged as strictly equivalent; errors at the boundary were concentrated on near-synonymy rather than completely unrelated categories.

\begin{table*}[t]
\centering
\caption{Generative classification performance (Precision, Recall, F1) under different prediction vs. ground-truth matching criteria ($\tau$ is semantic matching threshold). Absolute improvement over the trivial baseline aggregated over all product verticals is reported. 
\textbf{$\tau$-AUC} is the normalized trapezoidal area under the improvement-versus-threshold curve over $\tau\!\in\![0.70,0.90]$.}
\label{tab:threshold_sensitivity}
\resizebox{\textwidth}{!}{%
\begin{tabular}{lccccccccccccccc}
\toprule
\multirow{2}{*}{\textbf{Method}} & \multicolumn{3}{c}{\textbf{$\tau$=0.70}} & \multicolumn{3}{c}{\textbf{$\tau$=0.80}} & \multicolumn{3}{c}{\textbf{$\tau$=0.90}} & \multicolumn{3}{c}{\textbf{$\tau$=exact}} & \multicolumn{3}{c}{\textbf{$\tau$-AUC}} \\
\cmidrule(lr){2-4}\cmidrule(lr){5-7}\cmidrule(lr){8-10}\cmidrule(lr){11-13}\cmidrule(lr){14-16}
 & $\Delta$R & $\Delta$F1 & $\Delta$P & $\Delta$R & $\Delta$F1 & $\Delta$P & $\Delta$R & $\Delta$F1 & $\Delta$P & $\Delta$R & $\Delta$F1 & $\Delta$P & $\Delta$R & $\Delta$F1 & $\Delta$P \\
\midrule
Trivial & +0.0 & +0.0 & +0.0 & +0.0 & +0.0 & +0.0 & +0.0 & +0.0 & +0.0 & +0.0 & +0.0 & +0.0 & +0.0 & +0.0 & +0.0 \\
XLGen & +2.0 & +2.0 & +1.9 & +1.2 & +1.0 & +0.7 & +0.9 & +0.7 & +0.3 & +0.8 & +0.7 & +0.5 & +1.3 & +1.2 & +0.9 \\
GROOV & +1.8 & +2.3 & +3.1 & +1.2 & +1.5 & +1.9 & +1.0 & +1.2 & +1.5 & +1.1 & \textbf{+1.2} & \textbf{+1.5} & +1.3 & +1.6 & +2.1 \\
PatchRec & -0.1 & +0.8 & +2.5 & -0.7 & -0.2 & +0.8 & -0.8 & -0.4 & +0.4 & +0.1 & +0.4 & +1.1 & -0.6 & +0.0 & +1.1 \\
DEXML-DS & +1.1 & +2.0 & \textbf{+3.7} & +0.7 & +1.5 & \textbf{+2.9} & +0.7 & \textbf{+1.3} & \textbf{+2.6} & -0.1 & -0.1 & -0.1 & +0.8 & +1.6 & \textbf{+3.0} \\
DEXML-STk & -1.3 & -4.3 & -5.4 & -2.5 & -4.2 & -5.1 & -3.1 & -4.3 & -5.1 & -1.6 & -2.1 & -2.5 & -2.3 & -4.2 & -5.2 \\
\midrule
MERIT & \textbf{+14.1} & \textbf{+3.4} & -0.3 & \textbf{+10.3} & \textbf{+1.6} & -1.4 & \textbf{+9.1} & +1.0 & -1.7 & \textbf{+5.5} & +0.9 & -0.6 & \textbf{+11.0} & \textbf{+1.9} & -1.2 \\
\bottomrule
\end{tabular}}
\end{table*}

\section{Sensitivity to $k_g$}
\label{app:kg}
 
$k_g$ is the number of first-token hidden states from generated categories that are pooled into the propensity model's user representation (\S\ref{sec:dual-encoder}); the paper uses $k_g{=}5$. Table~\ref{tab:kg_ablation} sweeps $k_g \in \{1,3,5,7,10\}$ across all 20 product verticals and every Hit@$k$ cutoff, reporting the fraction of vertical--cutoff cells where MERIT beats both DEXML-DS and DEXML-STk simultaneously; every setting beats both baselines in 89--94\% of the 100 vertical--cutoff cells, so the choice is not fragile. Table~\ref{tab:kg_mean} reports the mean Hit@$k$ improvement over DEXML-STk for representative settings: $k_g{=}3$ is marginally strongest overall ($+15.0$), $k_g{=}5$ (used throughout the paper) is close behind ($+13.2$), and $k_g{=}10$ is somewhat lower ($+12.4$), all far ahead of the strongest baseline, DEXML-DS (mean gain $+5.2$).
 
\begin{table*}[h]
\centering
\caption{Sensitivity to $k_g$: fraction of the 20 product verticals, per Hit@$k$ cutoff, where the given $k_g$ beats both DEXML-DS and DEXML-STk simultaneously. \textbf{Overall} aggregates all 100 vertical--cutoff cells.}
\label{tab:kg_ablation}
\small
\begin{tabular}{@{}lcccccc@{}}
\toprule
$k_g$ & @0.5k & @1k & @5k & @10k & @20k & Overall \\
\midrule
1         & 16/20 & 17/20 & 18/20 & 20/20 & 20/20 & 91\% \\
3         & 18/20 & 18/20 & 19/20 & 19/20 & 20/20 & 94\% \\
5 (paper) & 17/20 & 17/20 & 19/20 & 19/20 & 19/20 & 91\% \\
7         & 16/20 & 16/20 & 19/20 & 19/20 & 19/20 & 89\% \\
10        & 17/20 & 17/20 & 18/20 & 19/20 & 19/20 & 90\% \\
\bottomrule
\end{tabular}
\end{table*}
 
\begin{table*}[h]
\centering
\caption{Mean Hit@$k$ improvement (absolute, averaged across all 20 verticals) over DEXML-STk, for representative $k_g$ settings; DEXML-DS shown for reference.}
\label{tab:kg_mean}
\small
\begin{tabular}{@{}lcccccc@{}}
\toprule
Method & @0.5k & @1k & @5k & @10k & @20k & Mean \\
\midrule
DEXML-DS & +5.5 & +5.3 & +4.7 & +5.3 & +4.9 & +5.2 \\
MERIT, $k_g{=}3$ & \textbf{+14.3} & \textbf{+12.0} & \textbf{+12.5} & \textbf{+16.4} & \textbf{+19.8} & \textbf{+15.0} \\
MERIT, $k_g{=}5$ (paper) & +11.6 & +10.3 & +11.1 & +14.9 & +18.2 & +13.2 \\
MERIT, $k_g{=}10$ & +11.4 & +10.0 & +10.5 & +13.9 & +16.3 & +12.4 \\
\bottomrule
\end{tabular}
\end{table*}

\begin{table*}[!t]
\centering
\caption{Extended generative-classification results, part A (absolute pp improvement over the trivial baseline). Companion to Table~\ref{tab:main_results}; baseline short-names are defined in \S\ref{sec:setup-baselines}.}
\label{tab:ext_genA}
\small
\setlength{\tabcolsep}{2pt}
\resizebox{\textwidth}{!}{
\begin{tabular}{@{}l@{\hspace{4pt}}c@{\hspace{4pt}}c@{\hspace{4pt}}c@{\hspace{4pt}}c@{\hspace{4pt}}c@{\hspace{4pt}}c@{\hspace{4pt}}c@{\hspace{4pt}}c@{\hspace{4pt}}c@{\hspace{4pt}}c@{\hspace{4pt}}c@{\hspace{4pt}}c@{\hspace{4pt}}c@{\hspace{4pt}}c@{\hspace{4pt}}c@{\hspace{4pt}}c@{\hspace{4pt}}c@{\hspace{4pt}}c@{\hspace{4pt}}c@{\hspace{4pt}}c@{\hspace{4pt}}c@{}}
\toprule
\multirow{2}{*}{\textbf{Method}} & \multicolumn{3}{c}{\textbf{Global}} & \multicolumn{3}{c}{\textbf{Beauty}} & \multicolumn{3}{c}{\textbf{Apparel}} & \multicolumn{3}{c}{\textbf{Wireless}} & \multicolumn{3}{c}{\textbf{Drugstore}} & \multicolumn{3}{c}{\textbf{Kitchen}} & \multicolumn{3}{c}{\textbf{Home}} \\
\cmidrule(lr){2-4} \cmidrule(lr){5-7} \cmidrule(lr){8-10} \cmidrule(lr){11-13} \cmidrule(lr){14-16} \cmidrule(lr){17-19} \cmidrule(lr){20-22}
 & R & F1 & P & R & F1 & P & R & F1 & P & R & F1 & P & R & F1 & P & R & F1 & P & R & F1 & P \\
\midrule
\multicolumn{22}{@{}l}{\textit{Baselines}} \\
XLGen & +2.1 & +2.0 & +1.9 & +2.8 & +2.2 & +1.4 & +3.5 & +4.0 & +4.7 & +6.9 & +5.8 & +4.4 & +4.3 & +5.0 & +5.8 & -3.3 & -3.6 & -3.9 & +0.9 & +0.7 & +0.5 \\
GROOV & +1.8 & +2.3 & +2.9 & +2.3 & +2.4 & +2.4 & +3.9 & \textbf{+5.3} & \textbf{+7.5} & +3.3 & +3.2 & +2.9 & +3.4 & +4.6 & \textbf{+6.4} & -3.5 & -3.4 & -3.2 & +1.8 & +2.0 & +2.2 \\
PatchRec & +0.2 & +1.2 & +3.1 & -2.3 & -2.3 & -2.1 & -0.2 & +1.2 & +5.1 & +4.5 & +5.1 & +5.9 & -2.3 & -1.9 & -1.0 & -4.3 & -3.4 & \textbf{-1.8} & +8.3 & +11.4 & \textbf{+17.0} \\
DEXML-DS & +1.2 & +2.2 & \textbf{+3.8} & +2.6 & +3.1 & \textbf{+3.7} & +1.8 & +2.6 & +3.9 & +11.5 & \textbf{+15.4} & \textbf{+22.6} & +1.8 & +2.7 & +4.0 & -3.8 & -3.7 & -3.6 & -0.8 & -0.6 & -0.3 \\
DEXML-STk & -0.9 & -4.1 & -5.2 & +4.1 & -3.1 & -5.7 & -1.8 & -4.2 & -5.1 & -5.3 & -6.1 & -7.2 & +1.3 & -1.9 & -2.9 & -4.2 & -7.6 & -8.6 & +1.1 & -1.9 & -2.7 \\
\midrule
\multicolumn{22}{@{}l}{\textit{Ablation: training objective (ours)}} \\
Standard Teacher Forcing & +0.2 & +0.5 & +1.0 & -0.6 & -0.9 & -1.4 & -0.0 & +0.8 & +2.5 & +4.0 & +3.8 & +3.3 & +1.4 & +2.2 & +3.4 & -4.1 & -3.8 & -3.5 & +3.1 & +3.5 & +4.2 \\
Permutation invariance & +3.0 & +1.6 & +0.4 & +4.3 & +1.2 & -1.3 & -0.7 & -0.9 & -1.0 & +13.5 & +7.9 & +4.0 & -0.2 & -0.4 & -0.7 & -0.3 & -1.1 & -1.9 & +12.3 & +10.5 & +8.9 \\
Cross-entropy Invariance & +3.4 & +1.9 & +0.6 & +3.9 & +0.6 & -2.0 & -0.4 & -0.7 & -1.2 & +15.9 & +9.4 & +4.9 & +0.7 & +0.2 & -0.2 & -0.5 & -1.3 & -2.1 & +12.4 & \textbf{+11.9} & +11.5 \\
\midrule
\textbf{MERIT} (1-beam) & \textbf{+14.1} & \textbf{+3.4} & -0.2 & \textbf{+15.5} & \textbf{+3.2} & -1.4 & \textbf{+13.8} & +4.5 & +0.9 & \textbf{+22.6} & +5.5 & +0.1 & \textbf{+15.9} & \textbf{+6.5} & +3.0 & \textbf{+11.4} & \textbf{-0.8} & -4.3 & \textbf{+13.7} & +3.7 & +0.9 \\
\bottomrule
\end{tabular}
}
\end{table*}

\begin{table*}[!t]
\centering
\caption{Extended generative-classification results, part B (absolute pp improvement over the trivial baseline). Companion to Table~\ref{tab:main_results}; baseline short-names are defined in \S\ref{sec:setup-baselines}.}
\label{tab:ext_genB}
\small
\setlength{\tabcolsep}{2pt}
\resizebox{\textwidth}{!}{
\begin{tabular}{@{}l@{\hspace{4pt}}c@{\hspace{4pt}}c@{\hspace{4pt}}c@{\hspace{4pt}}c@{\hspace{4pt}}c@{\hspace{4pt}}c@{\hspace{4pt}}c@{\hspace{4pt}}c@{\hspace{4pt}}c@{\hspace{4pt}}c@{\hspace{4pt}}c@{\hspace{4pt}}c@{\hspace{4pt}}c@{\hspace{4pt}}c@{\hspace{4pt}}c@{\hspace{4pt}}c@{\hspace{4pt}}c@{\hspace{4pt}}c@{\hspace{4pt}}c@{\hspace{4pt}}c@{\hspace{4pt}}c@{}}
\toprule
\multirow{2}{*}{\textbf{Method}} & \multicolumn{3}{c}{\textbf{PC}} & \multicolumn{3}{c}{\textbf{Grocery}} & \multicolumn{3}{c}{\textbf{Shoes}} & \multicolumn{3}{c}{\textbf{Book}} & \multicolumn{3}{c}{\textbf{Wireless Acc.}} & \multicolumn{3}{c}{\textbf{Electronics}} & \multicolumn{3}{c}{\textbf{Home Improv.}} \\
\cmidrule(lr){2-4} \cmidrule(lr){5-7} \cmidrule(lr){8-10} \cmidrule(lr){11-13} \cmidrule(lr){14-16} \cmidrule(lr){17-19} \cmidrule(lr){20-22}
 & R & F1 & P & R & F1 & P & R & F1 & P & R & F1 & P & R & F1 & P & R & F1 & P & R & F1 & P \\
\midrule
\multicolumn{22}{@{}l}{\textit{Baselines}} \\
XLGen & +0.8 & +0.6 & +0.2 & +2.7 & +1.8 & +0.3 & +1.3 & +1.3 & +1.1 & +0.3 & +0.2 & +0.1 & +1.4 & +1.1 & +0.8 & +1.1 & +0.9 & +0.5 & +1.4 & +1.3 & +1.3 \\
GROOV & +0.3 & +0.5 & +0.8 & +1.8 & +1.7 & \textbf{+1.3} & +2.2 & +3.1 & \textbf{+4.9} & +3.5 & \textbf{+4.1} & \textbf{+4.8} & +0.7 & +0.8 & +1.0 & +0.2 & +0.3 & +0.6 & +0.9 & +1.3 & +1.7 \\
PatchRec & -2.1 & -2.1 & -1.8 & -3.8 & -4.3 & -4.6 & -1.4 & -0.5 & +3.2 & -1.8 & -1.1 & +0.0 & -1.5 & -1.4 & -0.9 & -1.0 & -0.9 & -0.6 & +12.5 & \textbf{+15.3} & \textbf{+19.2} \\
DEXML-DS & +1.9 & \textbf{+3.7} & \textbf{+8.5} & -0.0 & -0.8 & -2.2 & -0.5 & -0.5 & -0.6 & -3.1 & -2.8 & -2.4 & +1.1 & +2.3 & \textbf{+5.0} & +1.6 & +2.6 & \textbf{+4.5} & +0.2 & +0.7 & +1.2 \\
DEXML-STk & -3.9 & -5.5 & -6.8 & -3.5 & -6.8 & -9.5 & +2.9 & -5.0 & -8.5 & -4.8 & -5.0 & -5.1 & -3.6 & -4.3 & -4.9 & -3.2 & -4.4 & -5.4 & +0.3 & -2.6 & -3.0 \\
\midrule
\multicolumn{22}{@{}l}{\textit{Ablation: training objective (ours)}} \\
Standard Teacher Forcing & -0.4 & -0.3 & -0.3 & -3.2 & -4.1 & -5.5 & -3.2 & -3.7 & -4.3 & +0.0 & +0.4 & +0.8 & -0.3 & -0.2 & +0.0 & -1.0 & -1.0 & -1.0 & +2.5 & +3.1 & +3.7 \\
Permutation invariance & +1.5 & -0.5 & -2.3 & -3.7 & -5.4 & -7.8 & -2.3 & -3.8 & -6.0 & +0.4 & -1.1 & -1.9 & +2.0 & +1.0 & +0.0 & +0.5 & -0.6 & -1.8 & +16.2 & +12.6 & +10.2 \\
Cross-entropy Invariance & +2.1 & -0.3 & -2.4 & -3.0 & -5.1 & -7.8 & -2.4 & -4.1 & -6.5 & +2.3 & -0.0 & -1.2 & +4.0 & +2.8 & +1.6 & +1.5 & +0.3 & -1.1 & +14.8 & +11.7 & +9.6 \\
\midrule
\textbf{MERIT} (1-beam) & \textbf{+13.2} & +2.7 & -1.4 & \textbf{+14.7} & \textbf{+2.8} & -3.0 & \textbf{+13.6} & \textbf{+3.6} & -2.1 & \textbf{+8.4} & +0.5 & -1.5 & \textbf{+12.9} & \textbf{+3.3} & +0.0 & \textbf{+11.5} & \textbf{+2.8} & -0.7 & \textbf{+16.8} & +4.2 & +1.2 \\
\bottomrule
\end{tabular}
}
\end{table*}

\begin{table*}[!t]
\centering
\caption{Extended generative-classification results, part C (absolute pp improvement over the trivial baseline). Companion to Table~\ref{tab:main_results}; baseline short-names are defined in \S\ref{sec:setup-baselines}.}
\label{tab:ext_genC}
\small
\setlength{\tabcolsep}{2pt}
\resizebox{\textwidth}{!}{
\begin{tabular}{@{}l@{\hspace{4pt}}c@{\hspace{4pt}}c@{\hspace{4pt}}c@{\hspace{4pt}}c@{\hspace{4pt}}c@{\hspace{4pt}}c@{\hspace{4pt}}c@{\hspace{4pt}}c@{\hspace{4pt}}c@{\hspace{4pt}}c@{\hspace{4pt}}c@{\hspace{4pt}}c@{\hspace{4pt}}c@{\hspace{4pt}}c@{\hspace{4pt}}c@{\hspace{4pt}}c@{\hspace{4pt}}c@{\hspace{4pt}}c@{\hspace{4pt}}c@{\hspace{4pt}}c@{\hspace{4pt}}c@{}}
\toprule
\multirow{2}{*}{\textbf{Method}} & \multicolumn{3}{c}{\textbf{Automotive}} & \multicolumn{3}{c}{\textbf{Baby Product}} & \multicolumn{3}{c}{\textbf{Sports}} & \multicolumn{3}{c}{\textbf{Toys}} & \multicolumn{3}{c}{\textbf{BISS}} & \multicolumn{3}{c}{\textbf{Lawn \& Garden}} & \multicolumn{3}{c}{\textbf{Personal Care}} \\
\cmidrule(lr){2-4} \cmidrule(lr){5-7} \cmidrule(lr){8-10} \cmidrule(lr){11-13} \cmidrule(lr){14-16} \cmidrule(lr){17-19} \cmidrule(lr){20-22}
 & R & F1 & P & R & F1 & P & R & F1 & P & R & F1 & P & R & F1 & P & R & F1 & P & R & F1 & P \\
\midrule
\multicolumn{22}{@{}l}{\textit{Baselines}} \\
XLGen & +1.1 & +1.0 & +0.9 & +5.3 & +5.9 & +6.6 & +1.7 & +1.6 & +1.4 & +0.5 & +0.4 & +0.3 & +0.9 & +0.8 & +0.7 & +2.2 & +2.3 & +2.5 & +2.1 & +1.8 & +1.5 \\
GROOV & +0.8 & +1.1 & +1.3 & -0.4 & +0.6 & +1.9 & +2.2 & +2.6 & \textbf{+3.3} & +2.0 & \textbf{+2.3} & \textbf{+2.7} & +0.6 & +0.9 & +1.3 & +0.5 & +1.0 & +1.7 & +3.1 & +3.5 & \textbf{+4.0} \\
PatchRec & +11.3 & \textbf{+14.3} & \textbf{+18.7} & -4.4 & -3.9 & -3.2 & -1.1 & -0.6 & +0.7 & -1.2 & -1.1 & -0.7 & +3.0 & +4.4 & \textbf{+6.1} & +4.3 & +6.0 & \textbf{+8.7} & -2.3 & -2.4 & -2.3 \\
DEXML-DS & +3.3 & +4.7 & +6.6 & +3.4 & \textbf{+6.1} & \textbf{+10.6} & -0.7 & -0.5 & -0.1 & -0.3 & -0.2 & -0.2 & -0.3 & +0.1 & +0.5 & +1.6 & +2.2 & +3.2 & +3.2 & +3.6 & \textbf{+4.0} \\
DEXML-STk & +1.1 & -2.3 & -2.9 & +7.7 & -3.0 & -5.2 & -0.9 & -4.4 & -6.0 & -1.5 & -2.2 & -2.5 & -1.6 & -3.4 & -3.7 & -1.3 & -3.4 & -4.0 & +0.3 & -2.3 & -3.1 \\
\midrule
\multicolumn{22}{@{}l}{\textit{Ablation: training objective (ours)}} \\
Standard Teacher Forcing & +1.1 & +1.3 & +1.6 & +2.0 & +3.2 & +5.0 & +1.3 & +1.7 & +2.4 & +0.3 & +0.4 & +0.6 & +0.6 & +1.1 & +1.7 & +3.1 & +3.8 & +4.7 & +0.2 & +0.3 & +0.4 \\
Permutation invariance & +2.5 & +1.3 & +0.6 & +7.4 & +5.9 & +4.7 & +2.0 & +0.6 & -0.8 & -0.2 & -0.6 & -1.0 & +9.4 & \textbf{+7.0} & +5.4 & +5.6 & +3.4 & +1.9 & -0.8 & -1.4 & -1.9 \\
Cross-entropy Invariance & +10.2 & +8.1 & +6.6 & -3.5 & -4.6 & -5.3 & +1.8 & +0.4 & -1.1 & -0.0 & -0.4 & -0.9 & +4.1 & +2.1 & +0.9 & +8.4 & \textbf{+6.3} & +4.7 & +0.1 & -0.6 & -1.2 \\
\midrule
\textbf{MERIT} (1-beam) & \textbf{+11.5} & +2.4 & +0.1 & \textbf{+14.6} & +3.1 & -0.8 & \textbf{+16.2} & \textbf{+5.1} & +0.5 & \textbf{+7.3} & +1.9 & +0.1 & \textbf{+12.2} & +2.5 & +0.1 & \textbf{+13.7} & +4.0 & +0.9 & \textbf{+14.7} & \textbf{+4.3} & +1.2 \\
\bottomrule
\end{tabular}
}
\end{table*}

\section{Full Per-Vertical Generative-Classification Results}
\label{app:full-gen}
Table~\ref{tab:main_results} reports five representative verticals; Tables~\ref{tab:ext_genA}-\ref{tab:ext_genC} give all 20 verticals plus the Global aggregate, split across three tables for readability, with Recall/F1/Precision kept together per vertical.

\section{Full Per-Vertical Propensity-Ranking Results}
\label{app:full-rank}
Table~\ref{tab:ranking} reports five verticals at three operating points; Table~\ref{tab:ext_ranking} gives all 20 verticals across every Hit@$k$ threshold, for both DEXML-DS and MERIT's 230M propensity model, each expressed relative to DEXML-STk. Averaged over the 20 verticals, MERIT improves Hit@5k by $+9.0$ points over DEXML-STk and $+4.3$ over DEXML-DS.

\begin{table*}[!t]
\centering
\caption{Extended propensity-ranking results (Hit@$k$, absolute pp improvement over the DEXML-STk baseline) across all 20 verticals and all $k$. Companion to Table~\ref{tab:ranking}; short-names are defined in \S\ref{sec:setup-baselines}. Both method blocks are expressed relative to DEXML-STk; MERIT uses the 230M propensity model on frozen generator embeddings.}
\label{tab:ext_ranking}
\scriptsize
\setlength{\tabcolsep}{4pt}
\resizebox{\textwidth}{!}{
\begin{tabular}{@{}l@{\hspace{6pt}}ccccc@{\hspace{10pt}}ccccc@{}}
\toprule
\multirow{2}{*}{\textbf{Vertical}} & \multicolumn{5}{c}{\textbf{DEXML-DS}} & \multicolumn{5}{c}{\textbf{MERIT (230M)}} \\
\cmidrule(lr){2-6} \cmidrule(lr){7-11}
 & @0.5k & @1k & @5k & @10k & @20k & @0.5k & @1k & @5k & @10k & @20k \\
\midrule
Beauty & +3.8 & +4.5 & +6.1 & +6.5 & +5.0 & \textbf{+20.8} & \textbf{+18.1} & \textbf{+12.6} & \textbf{+16.8} & \textbf{+21.3} \\
Apparel & -4.8 & -0.5 & +2.6 & +4.8 & +5.4 & \textbf{+1.4} & \textbf{+2.9} & \textbf{+9.3} & \textbf{+15.5} & \textbf{+19.7} \\
Wireless & \textbf{+64.6} & \textbf{+53.6} & \textbf{+22.1} & +25.3 & +22.5 & +58.4 & +51.5 & +21.5 & \textbf{+25.9} & \textbf{+27.7} \\
Drugstore & +3.2 & +2.4 & -0.5 & -1.0 & -1.5 & \textbf{+12.4} & \textbf{+13.0} & \textbf{+9.2} & \textbf{+12.6} & \textbf{+15.2} \\
Kitchen & -2.4 & -2.4 & -2.5 & -3.1 & -2.0 & \textbf{+0.0} & \textbf{+0.0} & \textbf{+0.6} & \textbf{+0.7} & \textbf{+3.1} \\
Home & -2.8 & -2.0 & -1.6 & -2.0 & -0.5 & \textbf{+6.0} & \textbf{+3.2} & \textbf{+5.2} & \textbf{+7.0} & \textbf{+12.2} \\
PC & -2.4 & -0.8 & +4.7 & +5.8 & +8.3 & \textbf{+6.0} & \textbf{+5.4} & \textbf{+6.2} & \textbf{+9.1} & \textbf{+13.6} \\
Grocery & \textbf{+18.2} & \textbf{+17.8} & +13.2 & +14.0 & +11.9 & +11.6 & +14.3 & \textbf{+21.0} & \textbf{+28.6} & \textbf{+28.2} \\
Shoes & -1.0 & -0.8 & +2.0 & +5.4 & +7.3 & \textbf{+17.0} & \textbf{+16.0} & \textbf{+12.8} & \textbf{+16.4} & \textbf{+21.2} \\
Book & -3.6 & -3.2 & -1.5 & -1.7 & -3.5 & \textbf{+16.4} & \textbf{+15.1} & \textbf{+14.3} & \textbf{+18.2} & \textbf{+20.2} \\
Wireless Acc. & +2.2 & +3.6 & +4.9 & +5.7 & +8.1 & \textbf{+4.8} & \textbf{+4.9} & \textbf{+6.3} & \textbf{+9.3} & \textbf{+15.3} \\
Electronics & -2.0 & -0.7 & +2.9 & +3.7 & +4.7 & \textbf{+3.4} & \textbf{+2.7} & \textbf{+7.1} & \textbf{+8.4} & \textbf{+14.0} \\
Home Improv. & +5.0 & +3.4 & +6.3 & +8.1 & +6.0 & \textbf{+7.4} & \textbf{+5.6} & \textbf{+10.1} & \textbf{+15.4} & \textbf{+19.6} \\
Automotive & \textbf{+2.0} & \textbf{+4.9} & \textbf{+7.5} & \textbf{+8.6} & +9.6 & +0.8 & +0.0 & +1.0 & +5.9 & \textbf{+15.9} \\
Baby Product & \textbf{+30.2} & \textbf{+23.1} & \textbf{+24.3} & \textbf{+23.7} & \textbf{+16.4} & -0.8 & +0.6 & +4.9 & +8.8 & +12.2 \\
Sports & -0.4 & -0.4 & -0.9 & +0.9 & +1.1 & \textbf{+3.0} & \textbf{+3.5} & \textbf{+7.8} & \textbf{+12.6} & \textbf{+15.8} \\
Toys & -0.8 & -0.7 & -3.3 & -4.7 & -3.5 & \textbf{+4.2} & \textbf{+3.2} & \textbf{+7.4} & \textbf{+10.3} & \textbf{+15.8} \\
BISS & \textbf{+1.0} & \textbf{+2.5} & \textbf{+4.6} & +3.6 & +2.0 & -0.6 & -0.4 & +2.3 & \textbf{+4.3} & \textbf{+8.6} \\
Lawn \& Garden & +0.4 & +1.4 & +6.1 & +7.5 & +6.7 & \textbf{+11.2} & \textbf{+9.1} & \textbf{+19.1} & \textbf{+24.6} & \textbf{+25.9} \\
Personal Care & +0.2 & +0.1 & -2.8 & -4.2 & -5.7 & \textbf{+2.2} & \textbf{+1.2} & \textbf{+1.9} & \textbf{+3.3} & \textbf{+8.1} \\
\bottomrule
\end{tabular}
}
\end{table*}

\section{Production A/B Test Details}
\label{app:abtest}
Table~\ref{tab:abtest} reports per-campaign A/B-test results, complementing the summary in \S\ref{sec:abtest}. Each campaign ran for one week, with users randomly assigned (50:50) to treatment (propensity-model-derived segments) or control (expert-configured segments). Treatment and control arms were matched on segment size, creative, and delivery, so segment quality is the only differing factor. Lift is evaluated with a Bayesian covariate-adjusted framework; the primary decision metric is the posterior probability of positive conversion lift, $P(\mathrm{lift}>0)$, rather than exclusion of zero from the 95\% credible interval. Our experimentation pipeline considers a treatment shippable when $P(\mathrm{lift}>0) \geq 0.90$, corresponding to 9-to-1 posterior odds in favor of positive impact. All three campaigns clear this pre-specified decision threshold. Two campaigns have 95\% credible intervals that overlap zero, while Campaign~3's interval lies entirely above zero; we therefore interpret the online results as positive Bayesian decision evidence across campaigns, with strongest individual evidence in Campaign~3.

\begin{table*}[t]
\centering
\caption{Sequence compression ablation. Results show absolute improvement over trivial baseline. Adaptive patching enables fitting full user histories within fixed context budgets.}
\label{tab:patching_ablation}
\small
\setlength{\tabcolsep}{3pt}
\resizebox{\textwidth}{!}{
\begin{tabular}{@{}l@{\hspace{4pt}}c@{\hspace{4pt}}c@{\hspace{4pt}}c@{\hspace{4pt}}c@{\hspace{4pt}}c@{\hspace{4pt}}c@{\hspace{4pt}}c@{\hspace{4pt}}c@{\hspace{4pt}}c@{\hspace{4pt}}c@{\hspace{4pt}}c@{\hspace{4pt}}c@{\hspace{4pt}}c@{\hspace{4pt}}c@{\hspace{4pt}}c@{\hspace{4pt}}c@{\hspace{4pt}}c@{\hspace{4pt}}c@{}}
\toprule
\multirow{2}{*}{\textbf{Context Strategy}} & \multicolumn{3}{c}{\textbf{Global}} & \multicolumn{3}{c}{\textbf{Beauty}} & \multicolumn{3}{c}{\textbf{Apparel}} & \multicolumn{3}{c}{\textbf{Electronics}} & \multicolumn{3}{c}{\textbf{Kitchen}} & \multicolumn{3}{c}{\textbf{Toys}} \\
\cmidrule(lr){2-4} \cmidrule(lr){5-7} \cmidrule(lr){8-10} \cmidrule(lr){11-13} \cmidrule(lr){14-16} \cmidrule(lr){17-19}
& R & F1 & P & R & F1 & P & R & F1 & P & R & F1 & P & R & F1 & P & R & F1 & P \\
\midrule
4k recent only & +13.6 & +3.5 & -0.1 & +14.9 & +3.3 & -1.3 & +13.2 & +4.5 & +1.0 & +11.3 & +2.9 & -0.6 & +11.0 & -0.7 & -4.1 & +7.0 & +2.0 & +0.2 \\
8k recent only & +13.9 & +3.5 & -0.2 & +15.4 & +3.4 & -1.3 & +13.6 & +4.5 & +0.9 & +11.2 & +2.7 & -0.7 & +11.1 & -0.8 & -4.2 & +7.2 & +1.9 & +0.2 \\
\textbf{8k full history (patched)} & \textbf{+14.1} & \textbf{+3.4} & -0.2 & \textbf{+15.5} & \textbf{+3.2} & -1.5 & \textbf{+13.8} & \textbf{+4.5} & +0.9 & \textbf{+11.5} & \textbf{+2.8} & -0.7 & \textbf{+11.4} & \textbf{-0.8} & -4.3 & \textbf{+7.3} & \textbf{+1.9} & +0.1 \\
\bottomrule
\end{tabular}
}
\end{table*}

\begin{table}[!h]
\centering
\caption{Per-campaign A/B-test results. Lift is the relative gain in
purchase conversions of treatment over control. CI denotes the 95\%
credible interval; $P(\mathrm{lift}{>}0)$ is the posterior probability of
positive lift under our Bayesian covariate-adjusted framework.}
\label{tab:abtest}
\small
\setlength{\tabcolsep}{4pt}
\begin{tabular}{@{}lccc@{}}
\toprule
\textbf{Campaign} & \textbf{Lift} & \textbf{95\% CI} & $P(\mathrm{lift}{>}0)$ \\
\midrule
Personal Care & $+0.23\%$ & $(-0.12\%,\,+0.57\%)$ & $0.90$ \\
Toys          & $+0.19\%$ & $(-0.06\%,\,+0.43\%)$ & $0.93$ \\
Sports        & $+0.37\%$ & $(+0.02\%,\,+0.72\%)$ & $0.98$ \\
\midrule
Average       & $+0.26\%$ & ---                   & ---    \\
\bottomrule
\end{tabular}
\end{table}

\section{Adaptive Patching Details}
\label{app:patching}

\paragraph{Compression strategy.} We apply non-overlapping window pooling
(window size $W=10$, stride $S=10$) to token embeddings in older
interaction windows, achieving $10\times$ compression in those segments.
This fits substantially more history, often the full year of
interactions, within the same context budget that would otherwise hold
only the most recent months. Unlike aggressive session-level compression,
this preserves finer temporal granularity while prioritizing recent
interactions, which remain uncompressed to retain the strongest
predictive signal.

\paragraph{Curriculum learning.} Training on patched sequences from the
start degrades performance, as the model must simultaneously learn the
prediction task and adapt to compressed representations. We therefore
introduce patching gradually over four phases: Phase~1 (0-25\%): full
sequences only; Phase~2 (25-50\%): patching applied with linearly
increasing probability $p(t)$; Phase~3 (50-75\%): patching applied
consistently to older interaction windows; Phase~4 (75-100\%): training
exclusively on patched sequences. This curriculum lets the model first
master the prediction task on uncompressed data before adapting to
compressed inputs.

\paragraph{Results.} Table~\ref{tab:patching_ablation} reports the sequence-compression ablation. Patching the full year of history into the $8$k budget yields a small but consistent recall gain over using only the most recent interactions ($+0.5$ global recall over $4$k-recent-only), confirming that the added history helps and that compression does not degrade the retained recent signal.

\section{Implementation Details and Hyperparameters}
\label{app:impl}

\paragraph{Generator.} MERIT uses \texttt{phi-3.5-mini} (3.8B) as the base
generator, fine-tuned with our exposure-mitigated objective and adaptive
patching. Following~\cite{sun2022clusterguided}, we manage the label space
via hierarchical generation: the 291k leaf categories are organized into
30k semantic clusters, and the model predicts clusters before leaf-level
categories.

\paragraph{Propensity model.} The propensity model is a 4-layer
transformer encoder followed by a 3-layer MLP (230M parameters total),
trained on the first-token hidden states ($k_g{=}5$) extracted from the
frozen generator. For categories we use pretrained
Cohere~\cite{CohereEmbed2023} embeddings; the propensity score is the MLP
output over the concatenated (user, category) representation.

\paragraph{Optimization.} We use AdamW (learning rate $2\times10^{-5}$),
batch size 32, and mixed precision (fp16) on $8\times$A100 GPUs. Generator
fine-tuning takes ${\sim}48$h; the propensity model takes ${\sim}3$h.

\paragraph{Hard-negative mining.} We draw 4 independent temperature samples
per training example (recall plateaus beyond 4). For the propensity model
we mine from the \emph{trained} mitigated generator and train against these
alongside popular and random negatives.

\paragraph{Evaluation.} For generative classification we report Recall,
Precision, and F1 with soft matching via Cohere cosine similarity
$\geq0.7$, calibrated on a small set of manually judged boundary pairs.
For propensity ranking we report Hit@$k$ when ranking users for an interest
cluster, aggregating leaf propensities to product-category granularity via the aggregation in section~\ref{sec:prop-results}.
All numbers are absolute percentage-point
improvements over a trivial baseline that predicts a user's 5 most
frequently purchased categories.

\end{document}